\documentclass[prd,twocolumn,showpacs,reprint,preprintnumbers,nofootinbib,superscriptaddress,aps]{revtex4-2}
\pdfoutput=1
\usepackage{graphicx}
\usepackage{ulem}
\usepackage{listings}
\usepackage{multirow,makecell,booktabs,siunitx}
\usepackage{amsfonts,amsmath,amssymb,bm,bbm,mathrsfs}
\usepackage{color}
\usepackage{enumitem}
\usepackage{physics}
\usepackage{url} 
\usepackage{fancybox,framed}
\usepackage[dvipsnames, usenames]{xcolor} 
\usepackage[nolist,nohyperlinks]{acronym}
\usepackage{comment}
\usepackage[multiple]{footmisc}

\newcommand{\D}{\text{d}}

\usepackage{hyperref}
\hypersetup{
    pdfnewwindow=true,      
    colorlinks=true,       
    linkcolor=Sepia,          
    citecolor=Sepia,        
    filecolor=Sepia,      
    urlcolor=Sepia        
}
\usepackage{cleveref}
\def\bi{\begin{itemize}[noitemsep,leftmargin=*]
\setlength\itemsep{1em}
        }
\def\ei{\end{itemize}}

\newcommand\be{\begin{equation}}
\newcommand\ee{\end{equation}}

\newcommand{\orcid}[1]{\begingroup
  \hypersetup{hidelinks}\href{https://orcid.org/#1}{\includegraphics[width=10pt]{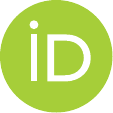}} \endgroup}

\begin{document}
\title{Importance of Shot Noise in the Search for an Isotropic Stochastic Gravitational-Wave Background with Next Generation Detectors}
\author{Haowen Zhong \orcid{0000-0001-8324-5158}\,}
\email{zhong461@umn.edu}
\affiliation{
 School of Physics and Astronomy, University of Minnesota, Minneapolis, Minnesota 55455, USA
}

\author{Vuk Mandic \orcid{0000-0001-6333-8621}\,}
\email{vuk@umn.edu}
\affiliation{
School of Physics and Astronomy, University of Minnesota, Minneapolis, Minnesota 55455, USA
}

\date{\today}
\begin{abstract}
We investigate the impact of shot noise on the stochastic gravitational wave background generated by binary neutron star mergers, and confirm that the overall background can be significantly influenced by relatively few neighboring, loud events. To mitigate the shot noise, we propose a procedure to remove nearby events by notching them out in the time-frequency domain. Additionally, we quantify the cosmic/sample variance of the resulting background after notching, and we study the deviation between the cross-correlation measurement and the theoretical prediction of the background. Taking both effects into account, we find that the resulting sensitivity loss in the search for an isotropic background formed by binary neutron star mergers is minimal, and is limited to $\lesssim 4\%$ below 40 Hz, and to $\lesssim 1\%$ above 40 Hz. 
\end{abstract}

\maketitle
\begin{acronym}
    \acro{GW}{gravitational wave}
    \acro{PSD}{power spectral density}
    \acro{GR}{general relativity}
    \acro{CBC}{compact binary coalescence}
    \acro{BH}{black hole}
    \acro{BBH}{binary black hole}
    \acro{BNS}{binary neutron stars}
    \acro{NSBH}{neutron star-black hole}
    \acro{SFR}{star formation rate}
    \acro{LVK}{LIGO-Virgo-KAGRA}
    \acro{ET}{Einstein Telescope}
    \acro{CE}{Cosmic Explorer}
    \acro{PE}{parameter estimation}
    \acro{SGWB}{stochastic gravitational wave background}
    \acro{XG}{next generation}
    \acro{SNR}{signal to noise ratio}
\end{acronym}
\section{Introduction}
Ten years after the first direct detection of the \ac{GW} event GW150914~\cite{LIGOScientific:2016aoc}, the observation of individually resolvable \ac{GW} events has become increasingly routine. By the end of the third observation (O3) run of the \ac{LVK} collaboration, over 90 \ac{GW} events have been detected~\cite{KAGRA:2021vkt} and a significant increase in this number is expected by the end of the fourth observation run (O4)~\cite{Mould:2022xeu, Callister:2024cdx}. However, due to the limited sensitivity of current-generation ground-based \ac{GW} detectors, including Advanced LIGO detectors at Hanford and Livingston~\cite{aLIGO}, Advanced Virgo~\cite{aVirgo, aVirgostatus} and KAGRA~\cite{KAGRA:2018plz}, we have not yet observed the \ac{SGWB}, which arises as an incoherent superposition of innumerable \ac{GW} signals across the whole history of the universe.

With ongoing enhancements in the sensitivity of ground-based GW observatories, prospects for detecting an astrophysical \ac{SGWB} from \ac{CBC} systems are highly promising in the near future~\cite{KAGRA:2021duu, KAGRA:2021kbb, Renzini:2022alw, O3stoch, Bellie:2023jlq}. Beyond the \ac{CBC}-generated \ac{SGWB}, several other sources of the \ac{SGWB} may exist, such as first-order phase transitions~\cite{PT1, PT2, PT3, PT4, PT5, PT6, PT7}, supernova explosions~\cite{Ferrari:1998ut, Buonanno:2004tp, Crocker:2015taa, Crocker:2017agi, Finkel:2021zgf}, standard inflationary models~\cite{Grishchuk:1974ny, Starobinsky:1979ty, Grishchuk:1993te}, axion inflation~\cite{Barnaby:2011qe}, and cosmic strings~\cite{Damour:2004kw, Siemens:2006yp, Olmez:2010bi, Regimbau:2011bm}. Detecting these (likely subdominant) \ac{SGWB} contributions would provide substantial scientific benefits. In particular, astrophysical \acp{SGWB} encode critical information about the nature and properties of their underlying sources~\cite{LIGOScientific:2020kqk, KAGRA:2021duu, Bavera:2021wmw}, while cosmological \acp{SGWB} could deliver unprecedented insights into the earliest epochs of the Universe, enabling us to probe fundamental physics at energy scales approaching the Planck scale~\cite{Grishchuk:1974ny, Starobinsky:1979ty, Grishchuk:1993te, Barnaby:2011qe, Damour:2004kw, Siemens:2006yp}.

The astrophysical \ac{SGWB} is expected to be dominated by \ac{CBC} signals, making it comparable to or even significantly stronger than the \ac{SGWB} of cosmological origin~\cite{Caldwell:2022qsj, Renzini:2022alw, Regimbau:2022mdu}. Extracting the cosmological background in the presence of this astrophysical foreground is therefore quite challenging and requires sophisticated statistical data analysis methods. Several Bayesian approaches have already been proposed, including the TBS method~\cite{TBS2}. Additionally, the Global-Fit approach~\cite{Littenberg:2020bxy, Littenberg:2023xpl} has been developed specifically for the future space-borne detector LISA~\cite{LISA:2017pwj}. These methods are computationally intensive, and their practical computational feasibility for real searches with the \ac{XG} \ac{GW} detectors remains to be fully demonstrated. In our previous works Ref.~\cite{Zhong:2022ylh,Zhong:2024dss, Zhong:2025qno}, we introduced a two-step solution that first removes the resolvable \ac{BBH} events
using a \textit{notching method}, and then simultaneously fits the \ac{BNS} population hyperparameters $\bm\Lambda_\mathrm{BNS}$ and the underlying cosmological SGWB, following the \textit{joint analysis} proposed in Ref.~\cite{Callister:2020arv}. As a result, we demonstrated--for the first time--the simultaneous inference of parameters characterizing both the dominant astrophysical foreground and a significantly weaker cosmological background.

\begin{figure*}
    \centering
    \includegraphics[width=0.95\linewidth]{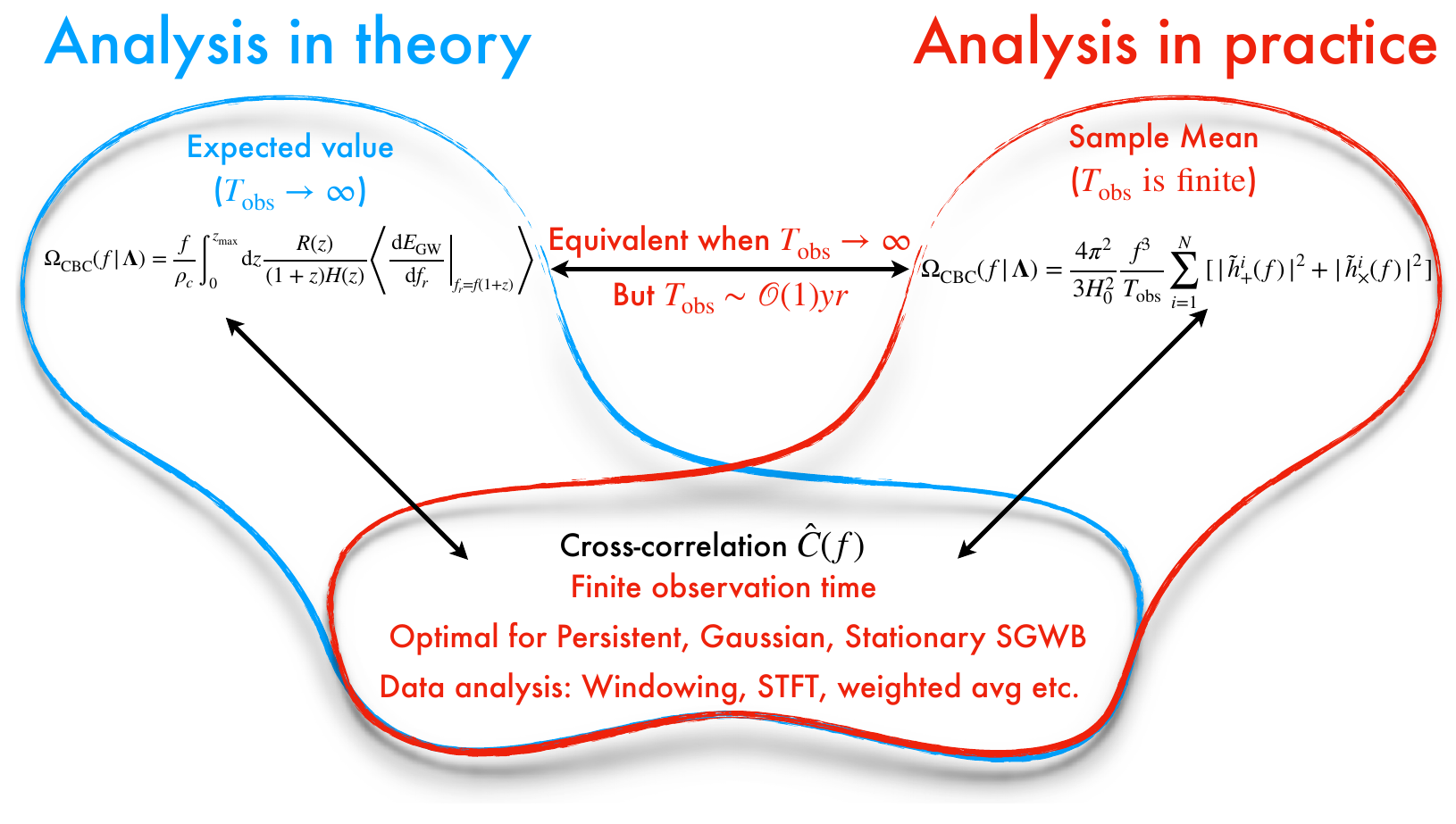}
    \caption{A schematic comparison between the theoretical analysis and the practical analysis of the \ac{CBC} foreground energy density. The blue bubble represents the “analysis in theory”, in which an theoretical expression for $\Omega_\mathrm{CBC}(f|\bm\Lambda)$ is shown. This integral has a rigorous one-to-one correspondence between population level hyperparameters $\bm\Lambda$ and $\Omega_\mathrm{CBC}(f|\bm\Lambda)$. The red bubble shows the “analysis in practice”, where the
    commonly used practical approximation is shown, which replaces the integral with a discrete summation over $N$ simulated events, hence depending on the exact simulation. Additionally, this approximation only holds when $T_\mathrm{obs}\to\infty$ or $N\to\infty$, whereas real observations are typically limited to a duration of $\mathcal{O}(1)$ year. The potential  fluctuation of the summation given different realizations of $N$ events must therefore be evaluated. At the bottom, we show the cross-correlation estimator $\widehat{C}(f)$, which measures the observed energy density spectrum of the \ac{SGWB}. The estimator $\widehat{C}(f)$ and corresponding uncertainty $\widehat{\sigma}(f)$ are defined in the Appendix~\ref{app: CC}, and we refer to Ref.~\cite{Romano:2016dpx, Allen:1997ad} for more details. Due to the factors we list below “Cross-correlation $\widehat{C}(f)$”, $\widehat{C}(f)$ and $\Omega_\mathrm{CBC}(f|\bm\Lambda)$ can potentially deviate from each other. Here STFT refers to the short-time Fourier transform.}
    \label{fig: Analysis}
\end{figure*}

However, we emphasize that many past studies (including our previous work Ref.~\cite{Zhong:2025qno}) treat the \ac{CBC} foreground as a persistent, Gaussian and stationary background. As we will discuss in detail here, this assumption does not accurately reflect the reality and, consequently, these analyses may underestimate the measurement uncertainties. Before proceeding with a detailed discussion, we emphasize that in this work we use the term \textit{foreground} to refer to the superposition of \ac{GW} signals from \textit{all} \ac{CBC} events, including those that are individually resolvable.

In Fig.~\ref{fig: Analysis}, we present a schematic comparison between theoretical analyses and practical analyses. Theoretically, one may model the \ac{CBC}-generated normalized SGWB energy density $\Omega_\mathrm{CBC}(f|\bm\Lambda)$ by an integral shown in the upper-left corner of Fig.~\ref{fig: Analysis}, where the equation is explicitly given by~\cite{Phinney:2001di,Regimbau:2022mdu,Callister:2020arv,Renzini:2022alw}
\begin{equation}
    \Omega_\mathrm{CBC}(f|\mathbf{\Lambda})=\frac{f}{\rho_c}\int_0^{z_{\mathrm{max}}}\D z\,\frac{R(z)}{(1+z)H(z)}\Bigg\langle\frac{\mathrm{d} E_{\mathrm{GW}}}{\mathrm{d} f_r}\Big|_{f_r=f(1+z)}\Bigg\rangle.
    \label{eq: integral}
\end{equation}
In the above expression, $f$ is the detector-frame frequency, $\bm{\Lambda}$ denotes the hyperparameters that determine the population details of the CBC system, $\rho_c=3H_0^2c^2/8\pi G$ is the critical energy density needed to close the universe, $z$ is the redshift, $R(z)$ is the source-frame merger rate of the corresponding CBC system, $H(z)$ is the Hubble parameter at the redshift $z$, $\frac{\D E_\mathrm{GW}}{\D f_r}$ is the energy emitted by the CBC system in the source frame at the source-frame frequency $f_r=f(1+z)$, and $\langle...\rangle$ denotes ensemble average. Notably, both $R(z)$ and $\left\langle\frac{\D E_\mathrm{GW}}{\D f_r}\right\rangle$ implicitly depend on the hyperparameters $\bm\Lambda$, and therefore for a given set of hyperparameters $\bm\Lambda$, $\Omega_\mathrm{CBC}(f|\bm\Lambda)$ is fully determined, establishing a clear mapping from hyperparameters $\bm\Lambda$ to the \ac{SGWB} energy density spectrum. However, the inverse mapping--from $\Omega_\mathrm{CBC}(f|\bm\Lambda)$ back to $\bm\Lambda$--may not be uniquely defined due to potential degeneracies among the hyperparameters.

In practice, one may approximate the integral by the summation shown in the upper-right corner of the Fig.~\ref{fig: Analysis}, given by
\begin{equation}
    \Omega_\mathrm{CBC}(f|\mathbf{\Lambda})=\frac{4\pi^2}{3H_0^2}\frac{f^3}{T_\mathrm{obs}}\sum_{i=1}^{N}[|\tilde{h}^i_+(f)|^2+|\tilde{h}^i_\times(f)|^2],
    \label{eq: summation}
\end{equation}
where $T_\mathrm{obs}$ denotes the observation time, $\tilde{h}^i_{+/\times}(f)$ denotes the frequency domain waveform of the $i$th CBC event for $+/\times$ polarization, and $N$ denotes the total number of events in a realization drawn from the population over an observation time $T_\mathrm{obs}$, accounting for both detectable and undetectable events. This approximation has been extensively employed in literature~\cite{Regimbau:2011rp,Regimbau:2012ir,Meacher:2014aca, Meacher:2015iua,Pan:2023naq,Li:2024iua}. Recently, Ref.~\cite{Belgacem:2024ohp} for the first time rigorously proved that this summation will finally converge to the integral when $T_\mathrm{obs}\to\infty$ or equivalently $N\to\infty$, which legitimates us to approximate the integral by this summation. However, we highlight that this result only holds when $T_\mathrm{obs}$ is large enough\footnote{In practice, only when the statistical fluctuation of the foreground itself is much weaker than the detector noise can one claim $T_\mathrm{obs}$ is large enough. Therefore, the threshold of the long enough observational time $T_\mathrm{th}$ is a detector-related quantity.}. In practical scenarios, where $T_\mathrm{obs}$ typically spans $\mathcal{O}(1)$~yr, statistical fluctuations in $\Omega_\mathrm{CBC}(f|\bm\Lambda)$, as computed by Eq.~\eqref{eq: summation}, across different realizations of the same underlying population model must  be accounted for. The fluctuation introduces an additional uncertainty of the search, distinct from and complementary to the detector noise. 

In the lower-middle corner of Fig.~\ref{fig: Analysis}, we depict the measured energy density $\widehat{C}(f)$ obtained via the cross-correlation method~\cite{Allen:1997ad,Romano:2016dpx} (A brief introduction can be found in Appendix~\ref{app: CC}). Cross-correlation is among the most commonly employed techniques for measuring the energy density of the \ac{SGWB} using ground-based detectors in practical searches~\cite{O3stoch}. This method was originally proposed and optimized for detecting a persistent, Gaussian and stationary \ac{SGWB}. In contrast, a \ac{CBC} foreground is known as non-Gaussian, non-stationary and “popcorn-like”. Furthermore, Eq.~\eqref{eq: summation} relies on frequency domain waveform $\tilde{h}(f)$ derived from the Fourier transform of the \textit{complete} time-domain waveform. In real \ac{SGWB} searches, the complete time series is typically segmented into shorter, 50\% overlapping segments, each windowed prior to applying a short-time Fourier transform~\cite{pygwb}. This data analysis process could introduce additional potential sources of deviation of the measured spectrum from the expected one. Moreover, detector noise fluctuates throughout the observation period, prompting the adoption of a \textit{weighted-average} rather than a \textit{simple average} to compute the narrow-band estimator $\widehat{C}(f)$. Conversely, both Eq.~\eqref{eq: integral} and Eq.~\eqref{eq: summation} inherently correspond to simple arithmetic averages of the \ac{CBC} energy density over the entire observation duration. Thus, this discrepancy could represent another practical source of deviation.

In this work we focus on the detection of a \ac{SGWB} in the \ac{XG} scenario, where $\gtrsim99\%$ \acp{BBH} are expected to be individually resolvable~\cite{Evans:2021gyd,Abac:2025saz}. Consequently, it may no longer be necessary to define a \ac{BBH} foreground $\Omega_\mathrm{BBH}(f|\bm\Lambda)$, as individual detections alone are informative enough to provide us with sufficient information about the entire population. In contrast, only around $50\%$ of \ac{BNS} events are expected to be resolvable~\cite{Evans:2021gyd,Abac:2025saz}. Hence, in this work we focus on \ac{BNS} systems and study the two new sources of uncertainty highlighted above: (1) the energy density spectrum estimated by Eq.~\eqref{eq: summation} can differ from the result given by Eq.~\eqref{eq: integral} due to the fact that we do not have access to infinite number of sources in a period of observation time, and (2) the measured energy density spectrum based on the cross-correlation method can differ from the result predicted by Eq.~\eqref{eq: summation} due to the subtleties we list above. We emphasize that the methodologies we proposed herein are also applicable to both \ac{BBH} and \ac{NSBH} systems. We also refer interested readers to a recent work by Giarda \textit{et al.}~\cite{Giarda:2025ouf}, which studies the uncertainty of the \ac{BBH} foreground and uses a machine learning based method to estimate the corresponding uncertainty rapidly.

The remainder of this work is organized as follows. In Sec.~\ref{sec: BNS_Foreground}, we describe the population model used in our simulations and present the corresponding results. Sec.~\ref{sec: Subtraction} discusses methods for mitigating shot noise from nearby, loud events, including both subtraction method and notching methods. We then compare the cross-correlation results with the theoretical estimates of the \ac{BNS} foreground in Sec.~\ref{sec: CC}. We discuss and draw conclusions in Sec.~\ref{sec: Dis_Conc}. 

We provide additional supporting figures in Appendix~\ref{app: extra_plots}. Appendix~\ref{app: Fisher} outlines the Fisher analysis used in this work. Appendix~\ref{app: CC} and Appendix~\ref{app: Notching} provide brief introductions to the cross-correlation and notching methods, respectively. In Appendix~\ref{app: 4svs192s}, we investigate how cross-correlation results vary with different segment lengths $T$ and frequency resolutions $\delta f$. Finally, in Appendix~\ref{app: weighted_avg}, we consider using estimated \ac{PSD}--rather than assuming them to be known--in the cross-correlation analysis.

\section{BNS Foreground}\label{sec: BNS_Foreground}
In this Section, we first introduce the population details of the \ac{BNS} systems we adopt for simulations, and then show the resulting $\Omega_\mathrm{BNS}(f)$ computed by Eq.~\eqref{eq: summation}. 

\subsection{\ac{BNS} Population Model}
In this work, we adopt the same population model used in our previous work (c.f. Ref.~\cite{Zhong:2022ylh}). Specifically, we assume that the \ac{BNS} masses are uniformly distributed within the range $[1,2]\,\rm{M_\odot}$~\cite{KAGRA:2021duu,Landry:2021hvl}. We assume zero spins and an isotropic distribution of the orbital orientation and sky position for all binaries. To obtain the redshift distribution, we convolve the \ac{SFR}~\cite{Finkel:2021zgf, Vangioni:2014axa}
\begin{equation}
    R_f(z)=\mathcal{N}\frac{ae^{b(z-z_p)}}{a-b+be^{a(z-z_p)}}
    \label{eq: SFR}
\end{equation}
with a power-law time delay distribution $p(t_d) \propto t_d^p$ with $p=-1$. The source-frame merger rate $R_m(z)$ then reads
\begin{equation}
R_m(z)=\int_{t_{\min}}^{t_{\max}}R_f(z_f)p(t_d)\D t_d\,,
\label{eq: Rm}
\end{equation}
where $z_f=z[t(z)-t_d]$ is the redshift at binary formation, we set $t_{\min}=20$ Myr, and $t_{\max}$ to be the Hubble time. We fix the parameters in the \ac{SFR} to $z_p=2.00$, $a=2.37$ and $b=1.80$~\cite{Finkel:2021zgf}, while $\mathcal{N}$ is a normalization factor chosen so that the local merger rate $\mathcal{R}_0$ is consistent with the \ac{LVK} results~\cite{KAGRA:2021duu}.
In particular, for \acp{BNS} we set $\mathcal{R}_0=320 \, \mathrm{Gpc}^{-3} \,\mathrm{yr}^{-1}$, identical to our past paper~\cite{Zhong:2025qno}.

Given the above setting, we then simulate 200 one-year long lists of \ac{BNS} events as 200 individual realizations of the same underlying population. For reference, there are  $\sim490,000$ events for each realization.

We highlight that this work aims to demonstrate the two new sources of uncertainty under the assumption of a particular underlying \ac{BNS} population. If the assumed population model is changed, the results presented here may differ accordingly. We refer interested readers to Ref.~\cite{popstock} for a detailed discussion of how uncertainties in the population properties themselves propagate into uncertainties in $\Omega_\mathrm{gw}(f)$.
\subsection{Simulation Results}
\begin{figure}
    \centering
    \includegraphics[width=0.9\linewidth]{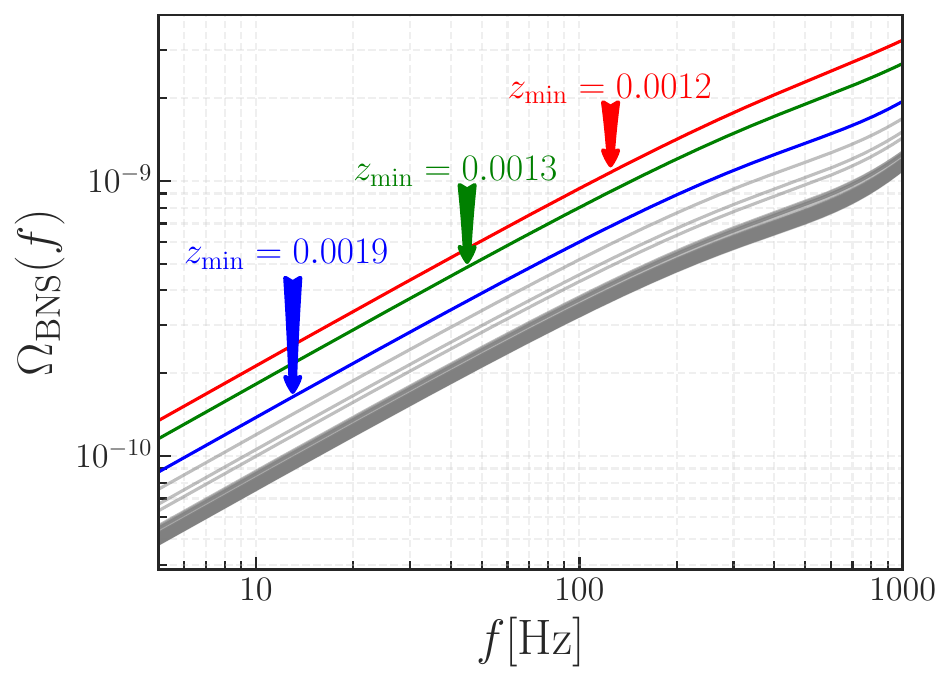}
    \caption{Energy density spectra $\Omega_\mathrm{BNS}(f)$ from two hundred simulated realizations with the identical underlying \ac{BNS} population. Each gray curve corresponds to a specific realization. The three realizations yielding the highest $\Omega_\mathrm{BNS}(f)$ are highlighted in red, green and blue.  For each of these, we also indicate the redshift $z_{\min}$ of the nearest event, which dominates the energy density in that realization. In these three realizations, the nearest event contributes roughly $65\%$(red), $59\%$(green) and $44\%$(blue) to the total $\Omega_\mathrm{BNS}(f)$.}
    \label{fig: BNS_Omega_all}
\end{figure}
With \ac{BNS} lists generated, we utilize Eq.~\eqref{eq: summation} to compute the corresponding $\Omega_\mathrm{BNS}(f)$ for each realization, where we uese the \texttt{Bilby}~\cite{Bilby} package to inject frequency domain waveform assuming \textsc{IMRPhenomXAS}~\cite{IMRXAS} waveform model, which includes only the dominant quadrupole mode. As pointed out by Ref.~\cite{popstock}, incorporating higher-order modes in the waveform has a negligible effect on the resulting $\Omega_\mathrm{gw}(f)$. We therefore do not consider the effect of higher-order modes in this work, and leave a more detailed investigation involving multiple waveform models to future studies. We present the results in Fig.~\ref{fig: BNS_Omega_all}, where each gray curve represents the computed $\Omega_\mathrm{BNS}(f)$ of a particular realization. Notably, several curves significantly deviate from the bulk distribution, exhibiting considerably higher energy densities. We highlight the three most prominent realizations using red, green and blue curves and indicate the redshift values of the nearest events in these particular cases. These realizations demonstrate that the resulting $\Omega_\mathrm{BNS}(f)$ can be dominated by nearby, exceptionally loud events, which is also known as the \textit{shot noise}. While shot noise has been extensively explored in the context of the \ac{SGWB} anisotropy in many studies, for instance, Ref.~\cite{Jenkins:2019nks, Jenkins:2019uzp, Alonso:2020mva, Yang:2023eqi}, its impact on the search for an isotropic \ac{SGWB} has yet to be thoroughly investigated. Although Ref.~\cite{Meacher:2014aca} addresses shot noise effects on the broadband estimator of $\Omega_\mathrm{CBC}(f_\mathrm{
ref})$ at a reference frequency, it does not consider its implications for the narrowband estimator of $\Omega_\mathrm{CBC}(f)$.

\begin{figure}[!htbp]
    \centering
    \includegraphics[width=0.9\linewidth]{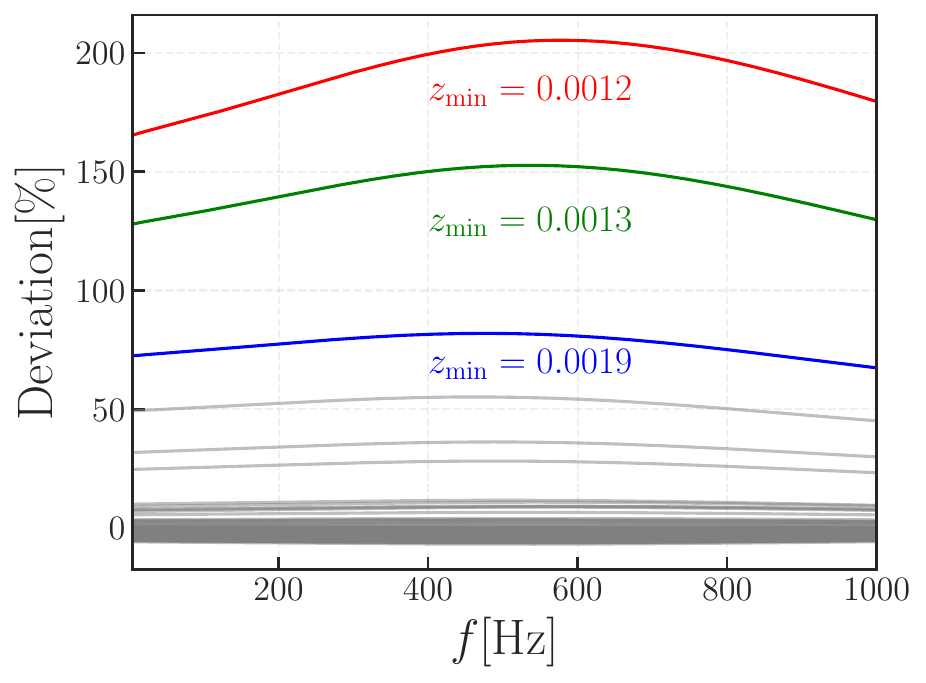}
    \caption{Similar to Fig.~\ref{fig: BNS_Omega_all}, but shows the relative deviation in percentage of each spectrum to the mean of two hundred realizations.}
    \label{fig: Deviation_all}
\end{figure}
In Fig.~\ref{fig: Deviation_all}, we illustrate the deviation of each calculated $\Omega_\mathrm{BNS}(f)$ from the mean $\Omega_\mathrm{BNS}(f)$ computed across all 200 realizations. Specifically, the deviation for the $i$th realization is defined as
\begin{equation}
    \mathrm{Deviation}_i(f):=\frac{\Omega_{\mathrm{BNS},i}(f)-\langle \Omega_\mathrm{BNS}(f)\rangle_{k=1}^{200}}{\langle \Omega_\mathrm{BNS}(f)\rangle_{k=1}^{200}}.
    \label{eq: Deviation}
\end{equation}

We also highlight three extreme realizations in red, green and blue as shown in Fig.~\ref{fig: BNS_Omega_all}. We observe that for the red curve, the deviation of $\Omega_\mathrm{BNS}(f)$ for this particular realization is $\gtrsim 150\%$ across the entire frequency band, significantly exceeding the bulk mean.

The above two figures suggest that $\Omega_\mathrm{BNS}(f)$ is \textit{not} a reliable quantity for a downstream analysis aimed at estimating the population hyperparameters $\bm\Lambda$. Since $\Omega_\mathrm{BNS}(f)$ for any individual realization could deviate from the bulk mean a lot as evidenced by Fig.~\ref{fig: Deviation_all}. Therefore, it can be problematic if one simply uses the mean of these 200 realizations (corresponding to approximately $\sim10^8$ events) as the theoretical prediction of $\Omega_\mathrm{BNS}(f|\bm{\Lambda})$ within a Bayesian framework. In practice, one may leverage Bayes' theorem to draw posterior samples of $\bm{\Lambda}$ given observed energy density spectrum $\widehat{C}(f)$
\begin{equation}
p(\bm{\Lambda}|\widehat{C}(f))\propto p(\bm{\Lambda})\mathscr{L}(\widehat{C}(f)|\bm\Lambda),
\end{equation}
where $p(\bm{\Lambda})$ is the prior distribution of hyperparameters, and $\mathscr{L}(\widehat{C}(f)|\bm{\Lambda})$ is the likelihood of observing the energy density spectrum $\widehat{C}(f)$ given $\bm{\Lambda}$. A Gaussian likelihood is often adopted in this context
\begin{equation}
    \mathscr{L}(\widehat{C}(f)|\bm{\Lambda})=\frac{1}{\sqrt{2\pi}\sigma_n(f)}\exp\left(-\frac{(\widehat{C}(f)-\Omega_\mathrm{BNS}^\mathrm{Model}(f|\bm\Lambda))^2}{2\sigma_n^2(f)}\right),
    \label{eq: likelihood_old}
\end{equation}
where $\sigma_n(f)$ is the corresponding uncertainty associated with $\widehat{C}(f)$.
However, we now have shown that this expression could be problematic due to three reasons: (1) It is not proper to simply take the ensemble average $\langle \Omega_\mathrm{BNS}(f)\rangle$ as the definition of the quantity $\Omega_\mathrm{BNS}^\mathrm{Model}(f|\bm\Lambda)$; (2) The above expression only considers the detector noise $\sigma_n(f)$, but ignores the statistical fluctuation of the \ac{BNS} foreground itself considering the finite observational time;\footnote{For the current generation of gravitational-wave detectors, due to the limited sensitivity, the statistical fluctuation of the foreground is expected to be weaker than the detector noise, so it is still safe to ignore this term, and still assume a Gaussian likelihood. This claim can be justified by Fig.~\ref{fig: Budget} and by noting that the noise PSD of current detectors is approximately an order of magnitude larger than that of a CE detector, which in turn leads to an approximately two-order-of-magnitude increase in $\sigma(f)$.}\textsuperscript{,}\footnote{If one does not aim to infer the population hyperparameters $\bm{\Lambda}$ of the corresponding foreground, but simply regard the foreground as a special realization of an ideal \ac{SGWB}, which is persistent, stationary and Gaussian, then one does not need to consider the statistical fluctuation.}(3) The deviations we show in Fig.~\ref{fig: Deviation_all} are not Gaussian, and hence the Gaussian likelihood itself may not hold in general.
\begin{figure*}[!htbp]
    \centering
    \includegraphics[width=0.9\linewidth]{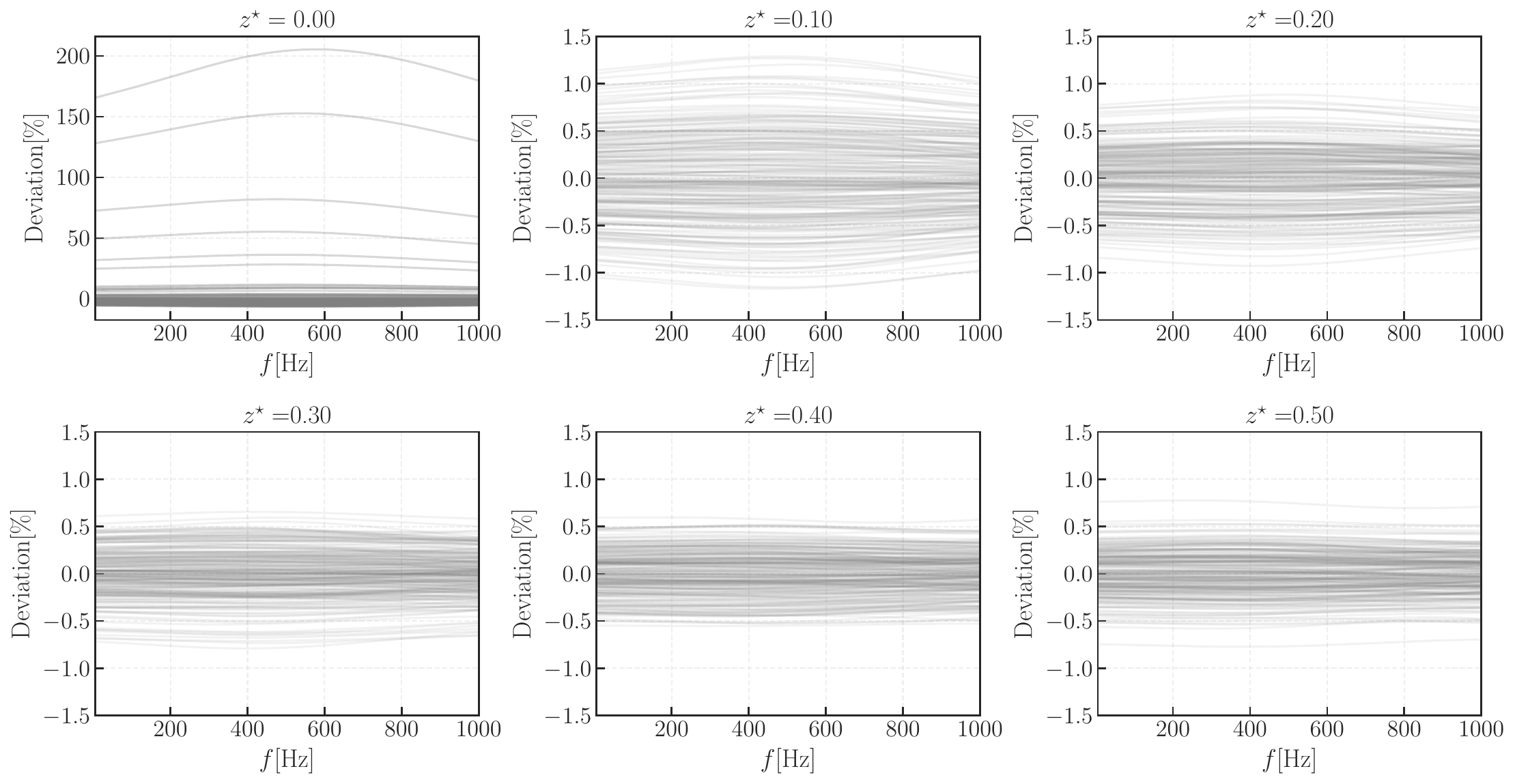}
    \caption{Similar to Fig.~\ref{fig: Deviation_all}, but now each here subplot corresponds to a different threshold $z^\star$ value. We only show six $z^\star$ values in this plot and the complete version is shown in Fig.~\ref{fig: Deviation_BNS_z_star}. As $z^\star$ increases, progressively more nearby events are removed from each realization. As a result, the outliers that appears prominently in the $z^\star=0.00$ case gradually disappear, and the overall deviation of each realization from the mean of all two hundred realizations systematically decreases with increasing $z^\star$.}
    \label{fig: Deviation_BNS_z_star_six}
\end{figure*}
\begin{figure*}[!htbp]
    \centering
    \includegraphics[width=0.9\linewidth]{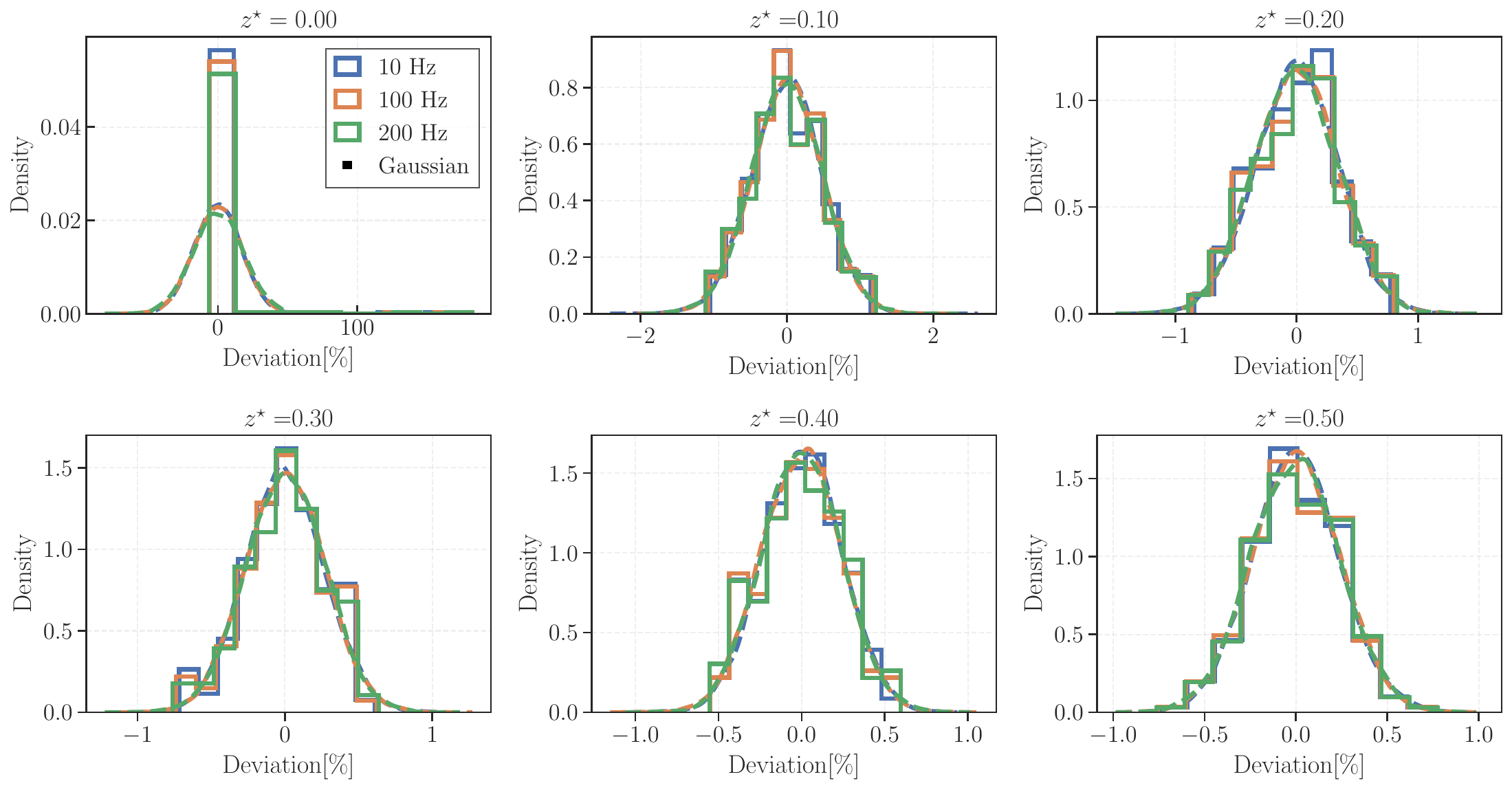}
    \caption{We aggregate all two hundred realizations from the previous plot to examine the distribution of deviations. We focus on three particular frequencies: 10 Hz, 100 Hz, 200 Hz. The results show that the deviation is largely frequency-independent. We also fit a Gaussian to each of the three frequencies for each subplot in dashed lines, and find that after removing nearby \ac{BNS} events, the deviations are well characterized by a simple Gaussian, suggesting that the corresponding $\Omega_\mathrm{BNS}^{z^\star}(f)$ also \textit{approximately} follows a Gaussian distribution. In the upper-left panel, Gaussian curves poorly match the histograms because the distributions have a pronounced long tail extending beyond $100\%$ (also see Fig.~\ref{fig: Deviation_all}). See Fig.~\ref{fig: Deviation_BNS_z_star_3freqs} for the complete version.}
    \label{fig: Deviation_BNS_z_star_3freqs_six}
\end{figure*}

In the remainder of this paper, we will discuss how to define a more robust quantity than $\Omega_\mathrm{BNS}(f)$ for the down-stream Bayesian analysis and modify the above likelihood accordingly to account for new sources of uncertainties.

\section{Suppressing the Shot Noise by Removing Nearby Events}\label{sec: Subtraction}
Per Fig.~\ref{fig: BNS_Omega_all} and Fig.~\ref{fig: Deviation_all}, we can already see that even a few neighboring events can dominate the whole \ac{BNS} background. In order to avoid the impact of these neighbors, we can remove them from the data and only focus on the remaining events. Therefore, we define a new quantity $\Omega_\mathrm{BNS}^{z^\star}(f)$ as following
\begin{equation}
\Omega_\mathrm{BNS}^{z^\star}(f)=\frac{f}{\rho_c}\int_{z^\star}^{z_{\mathrm{max}}}\mathrm{d} z\frac{R(z)}{(1+z)H(z)}\Bigg\langle\frac{\mathrm{d} E_{\mathrm{GW}}}{\mathrm{d} f_r}\Big|_{f_r=f(1+z)}\Bigg\rangle.
    \label{eq: integral_zstar}
\end{equation}
To simplify the notation, we suppress the dependence of the \ac{BNS} background on $\bm{\Lambda}$. Comparing the above definition and Eq.~\eqref{eq: integral}, we solely change the lower limit of the integral from 0 to $z^\star$\footnote{The similar cutoff of the lower bound in redshift has also been proposed in the studies of the shot noise for \ac{SGWB} anisotropies in Ref.~\cite{Jenkins:2019uzp, Alonso:2020mva}, where the cutoff is chosen to avoid the divergence of an integral.}. The corresponding summation version of $\Omega_\mathrm{BNS}^{z^\star}(f)$ reads
\begin{equation}
    \Omega_\mathrm{BNS}^{z^\star}(f)=\frac{4\pi^2}{3H_0^2}\frac{f^3}{T}\sum_{i=1}^{N}[|\tilde{h}^i_+(f)|^2+|\tilde{h}^i_\times(f)|^2]\Theta(z_i-z^\star),
    \label{eq: summation_zstar}
\end{equation}
where $\Theta(x)$ is the well-known Heaviside step function that
\begin{equation}
    \Theta(x):=\left\{\begin{aligned}
         &0,~x< 0 \\
         &1, ~x\geqslant0.
    \end{aligned}
    \right.
\end{equation}

In this work, we consider $z^\star$ values ranging from 0.01 to 0.50 in steps of 0.01. Based on Eq.~\eqref{eq: summation_zstar}, we compute $\Omega_\mathrm{BNS}^{z^\star}(f)$ for each $z^\star$ value, and evaluate the relative deviation of each realization from the mean over all realizations according to Eq.~\eqref{eq: Deviation}. We present the results in Fig.~\ref{fig: Deviation_BNS_z_star_six} and Fig.~\ref{fig: Deviation_BNS_z_star}, where each subplot corresponds to one specific $z^\star$ value. After removing nearby events, $\Omega_\mathrm{BNS}^{z^\star}(f)$ becomes significantly better behaved, i.e. outliers are eliminated, and the statistical fluctuation keeps decreasing as $z^\star$ increases. 

In Fig.~\ref{fig: Deviation_BNS_z_star_3freqs_six}, we focus on three particular frequencies: 10 Hz, 100 Hz and 200 Hz. For each $z^\star$ value we aggregate the relative deviations at these frequencies across all 200 realizations and plot the resulting distribution. Additionally, we fit a Gaussian to each of these three frequencies for comparison. We find that the deviation is not sensitive to frequency for $z^\star>0$ cases and that the overall distribution can be well approximated by a Gaussian if the nearby events are removed. Since we only simulate two hundred one-year-long realizations, the resulting distributions look not exactly the same as Gaussian distributions, whereas as the number of realizations and $T_\mathrm{obs}$ increase, the deviations are expected to converge to a Gaussian. A brief justification is as follows: if the observation time $T_\mathrm{obs}\to\infty$, and the number of \ac{BNS} events $N\to \infty$, then $\Omega_\mathrm{BNS}^{z^\star}(f)\approx N\langle \Omega_\mathrm{BNS}^{z^\star}(f)\rangle_{\mathrm{one~event}}$, where $\langle \Omega_\mathrm{BNS}^{z^\star}(f)\rangle_{\mathrm{one~event}}$ denotes the expected value of the energy density from a single \ac{BNS} event. In this regime and assuming a particular population, the observed $\Omega_\mathrm{BNS}^{z^\star}(f)$ solely depends on $N$, which follows a Poisson distribution, i.e. $N\sim\mathrm{Poisson}[\lambda]$, where $\lambda$ is the expected value of $N$ given $T_\mathrm{obs}$. For large $\lambda$, this Poisson distribution approaches a Gaussian, $N\sim\mathcal{N}(\lambda, \lambda)$, and so does the fluctuations in $\Omega_\mathrm{BNS}^{z^\star}(f)$.

\begin{figure*}[!htbp]
    \centering
    \includegraphics[width=1.0\linewidth]{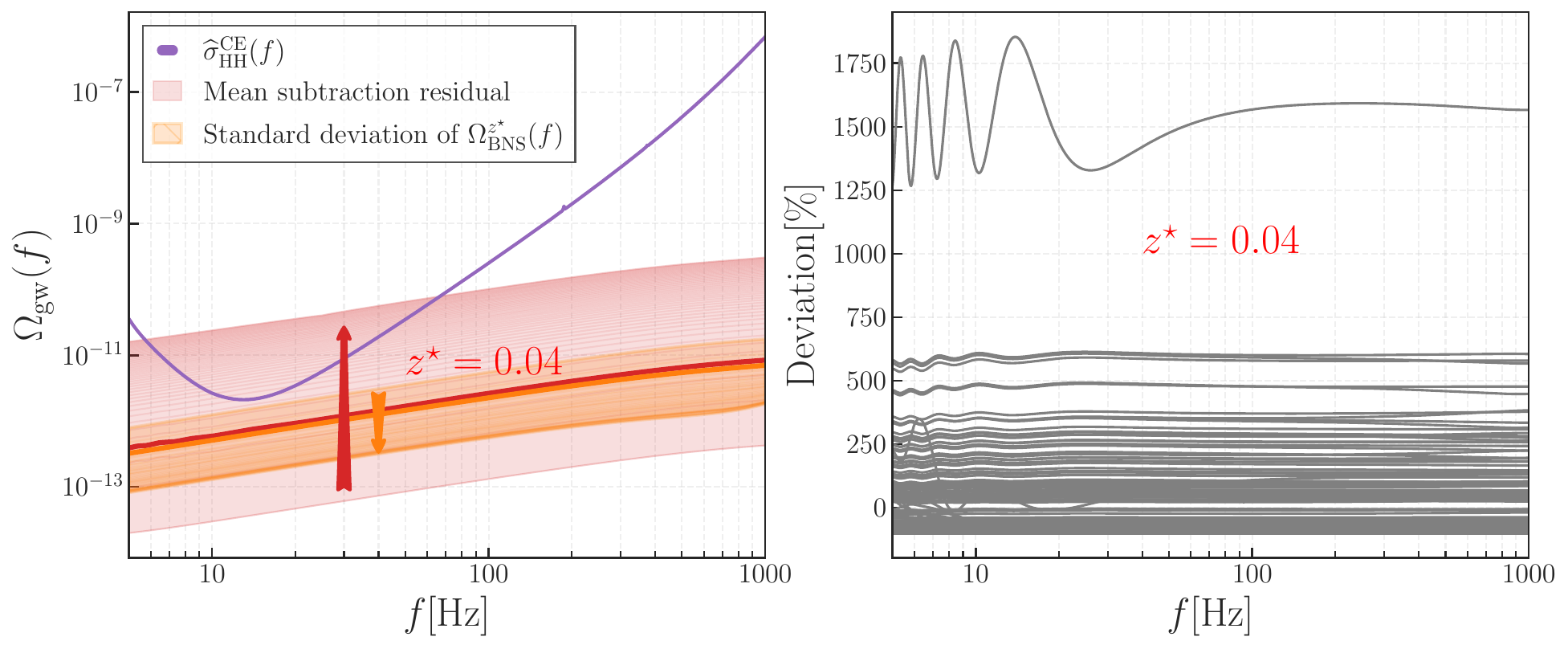}
    \caption{Left: Comparison between the mean subtraction residual foreground (red) and the standard deviation of $\Omega_\mathrm{BNS}^{z^\star}(f)$ (orange). Two arrows indicate how these quantities evolve with increasing $z^\star$. We highlight the case $z^\star=0.04$, where the mean subtraction residual begins to exceed the intrinsic fluctuation of the foreground. For reference, the uncertainty of a search for an isotropic \ac{SGWB} given one year observation time is shown in purple assuming a detector network consisting of two coincident and co-located detectors at LIGO Hanford site with \ac{CE}2 sensitivity~\cite{CE2_PSD}. Right: The deviation of the subtraction residual for each realization from the mean subtraction residual for $z^\star=0.04$ case. This indicates that the subtraction residual is also dominated by the residual of nearby loud events.}
    \label{fig: subtraction_vs_std}
\end{figure*}
We note that although in the definition of $\Omega_\mathrm{BNS}^{z^\star}(f)$ (c.f. Eq.~\eqref{eq: summation_zstar}) we only consider events with redshift greater than $z^\star$, in practice we can only remove those resolvable ones (i.e. those with sufficiently high signal-to-noise ratio), and those unresolvable ones form an unresolvable \ac{BNS} foreground. As a reminder, in this work the term “unresolvable foreground” refers to the foreground formed by those nearby but unresolvable signals, while in other studies, authors may use this term to refer to the foreground generated by all unresolvable events. We define the unresolvable foreground as below:
\begin{equation}
\begin{aligned}
\Omega_\mathrm{unres}^{z^\star}(f)=\frac{4\pi^2}{3H_0^2}\frac{f^3}{T}&\sum_{i=1}^{N}[|\tilde{h}^i_+(f)|^2+|\tilde{h}^i_\times(f)|^2]\\
&\times \Theta(z^\star-z_i)\Theta(\rho_\mathrm{th}-\rho_i),
    \label{eq: unres}
\end{aligned}
\end{equation}
where $\rho_\mathrm{th}=12$ is the \ac{SNR} threshold for detection and $\rho_i$ is the network \ac{SNR} of the $i$th event, which is introduced in  Appendix~\ref{app: Fisher}. 

In this work we consider two detector network configurations for different purposes. For detecting individually resolvable \ac{BNS} events, we use a network of two \ac{CE} detectors~\cite{CE, CE2_PSD} with 40 km arms at LIGO Hanford and LIGO Livingston, together with an \ac{ET} detector~\cite{ET,Abac:2025saz} at the Virgo site, matching the choice we made in the previous work (please see Ref.~\cite{Zhong:2025qno}). For the cross-correlation analysis, we instead consider a simpler network consisting of two co-located and co-aligned detectors at LIGO Hanford site operating at \ac{CE}2 sensitivity~\cite{CE, CE2_PSD}. We exclude an \ac{ET} detector here because, as shown in our earlier study Ref.~\cite{Zhong:2024dss}, adding it yields only a $\sim10\%$ improvement in overall sensitivity, whereas the computational cost is increased by an order of magnitude\footnote{There are ten baselines for the \{CE-CE-ET1-ET2-ET3\} network in total, but only a single baseline for the \{CE-CE\} network. Here we consider that one ET is composed of three colocated and independent detectors, we denote them
by ET$i$ ($i = 1, 2, 3$).}. 
Placing both detectors at the same site ensures the overlap reduction function $\gamma_\mathrm{HH}(f)\equiv 1$~\cite{Allen:1997ad, Romano:2016dpx} for all frequencies, enabling a seamless comparison between the energy density spectrum $\widehat{C}(f)$ estimated via the cross-correlation method and the theoretical summation $\Omega_\mathrm{BNS}^{z^\star}(f)$ (c.f. Sec.~\ref{sec: CC}). We emphasize that adopting this network for cross-correlation analysis yields a \textit{conservative} estimation of the sensitivity reduction due to three factors we will discuss below in Sec.~\ref{sec: CC}. Thus, this choice does not weaken our conclusions. Please see the first remark at the end of Sec.~\ref{sec: CC} for more information.

We stress that Eq.~\eqref{eq: unres} is an \textit{ad hoc} definition of the energy density contributed by these unresolvable events. As we discussed above, this summation is only valid when the event number $N$ is large enough per Ref.~\cite{Belgacem:2024ohp}, but the number of unresolvable \ac{BNS} events at redshifts $z\leqslant 0.5$ is of the order of $\mathcal{O}(100)$ for the \ac{XG} detector network we assume here. Nevertheless, Eq.~\eqref{eq: unres} can still serve as a rough estimation of the unresolvable foreground.

Literature considers two ways to remove nearby events. We will first consider directly subtracting the BNS signals in the frequency domain~\cite{Zhou:2022nmt,Zhou:2022otw} and we will point out the possible issues arising with this approach. Then, in the remaining part of the paper, we will discuss an alternative approach to remove events in the time-frequency domain using the notching method we proposed in Ref.~\cite{Zhong:2022ylh,Zhong:2024dss, Zhong:2025qno}.

Starting with the BNS subtraction approach, we note that the subtraction is never perfect and results in a residual \ac{BNS} foreground. Therefore, we define a subtraction residual foreground following the above formalism:
\begin{equation}
\begin{aligned}
&\Omega^{z^\star}_{\mathrm{resi}}(f)=\frac{4\pi^2}{3H_0^2}\frac{f^3}{T}\sum_{i=1}^N\Big[\left|\tilde{h}_+^i(\bm{\theta}^i_\mathrm{tr};f)-\tilde{h}_+^i(\bm{\theta}^i_\mathrm{rec};f)\right|^2\\
&+\left|\tilde{h}_\times^i(\bm{\theta}^i_\mathrm{tr};f)-\tilde{h}_\times^i(\bm{\theta}^i_\mathrm{rec};f)\right|^2\Big]\Theta(z^\star-z_i)\Theta(\rho_i-\rho_\mathrm{th}).
\end{aligned}
\end{equation}
In the above equation, $\bm{\theta}_\mathrm{tr}^i$ denotes the true parameters of the $i$th \ac{BNS} event, and $\bm{\theta}^i_\mathrm{rec}$ denotes the recovered parameters for the same event. We perform a Fisher analysis to determine $\bm{\theta}_\mathrm{rec}^i$, as described in the Appendix~\ref{app: Fisher}. We also refer to Ref.~\cite{Zhou:2022nmt, Zhou:2022otw, Zhong:2025qno} for more details.

In the left panel of Fig.~\ref{fig: subtraction_vs_std}, we compare the subtraction residual and the statistical fluctuation of the foreground, where the
red curves correspond to the mean of the subtraction residual across the 200 realizations for different $z^\star$ values, and the orange curves show the standard deviation of $\Omega_\mathrm{BNS}^{z^\star}(f)$, which is denoted by $\sigma_{\Omega_\mathrm{BNS}^{z^\star}}(f)$ in the later discussion. The orientation of the arrow shows how red/orange curves change as $z^\star$ increases. As $z^\star$ increases starting from zero, the mean subtraction residual increases rapidly and surpasses the standard deviation of $\Omega_\mathrm{BNS}^{z^\star}(f)$ when $z^\star=0.04$. After this point, the mean subtraction residual plateaus as $z^\star$ approaches 0.50. 
\begin{figure*}[!htbp]
    \centering
    \includegraphics[width=0.9\linewidth]{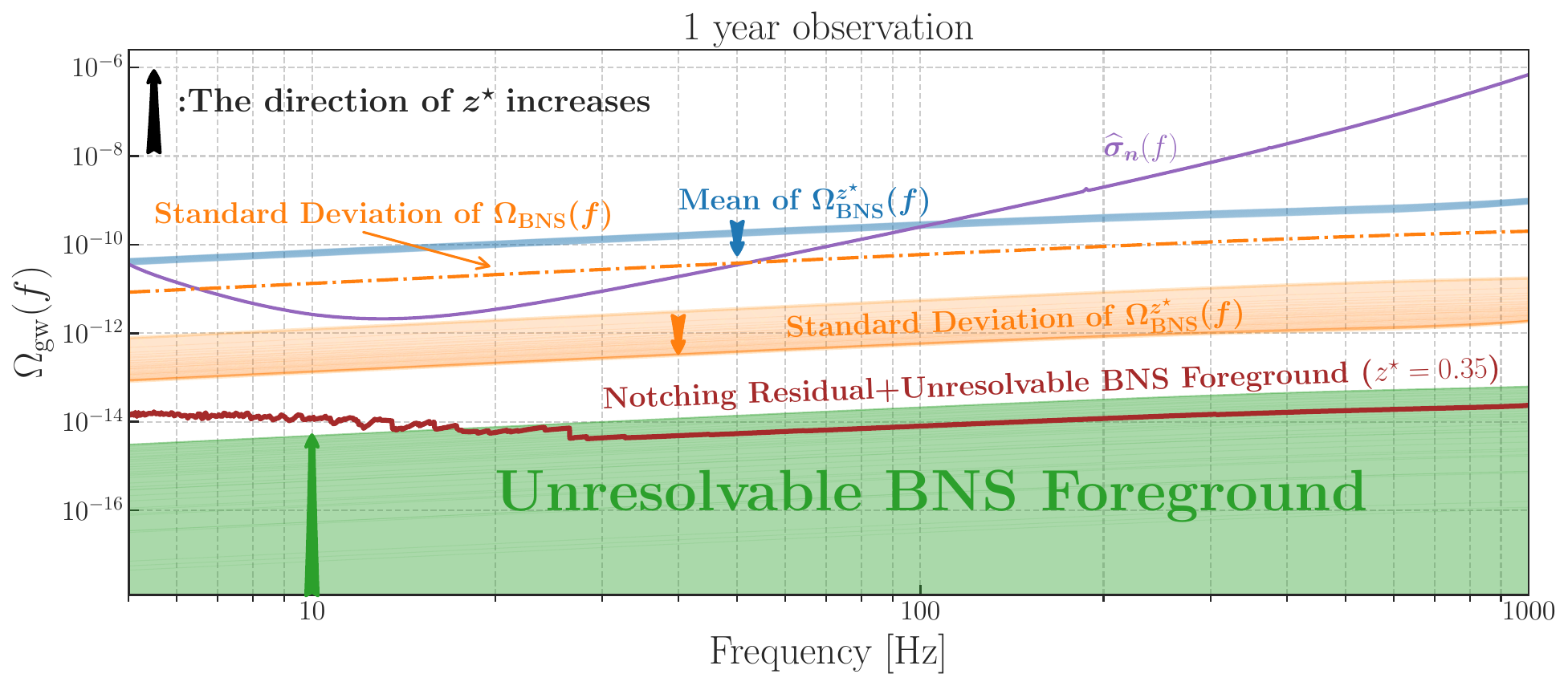}
    \caption{Energy density spectrum budget for four foreground components and the cross-correlation uncertainty, assuming one-year of observation. The blue curves correspond to the expected magnitudes of the \ac{BNS} foreground after removing nearby events, the orange ones show the statistical fluctuation of the foreground itself, and the green ones show the unresolvable foreground. The purple curve is identical to the one shown in Fig.~\ref{fig: subtraction_vs_std}. The orientation of the arrow indicates the trend of how these curves vary with increasing $z^\star$. We also plot the estimated notching residual plus the unresolvable foreground for the special case $z^\star=0.35$ in brown; see the main text for a detailed discussion.}
    \label{fig: Budget}
\end{figure*}

For reference, we also show $\widehat{\sigma}^{\mathrm{CE}}_\mathrm{HH}(f)$ in purple, which is the uncertainty of the cross-correlation spectrum $\widehat{C}(f)$ (c.f. Eq.~\ref{eq:stoch}) assuming one-year observation and two co-located and coincident detectors at LIGO Hanford site, both operating with \ac{CE}2 sensitivity~\cite{CE2_PSD}. The segment duration of the cross-correlation analysis is chosen to be $T=192$ s and $\delta f=1/32$ Hz. We refer to Appendix~\ref{app: CC} and Ref.~\cite{pygwb,Romano:2016dpx,Allen:1997ad} for more technical details regarding the cross-correlation analysis. For simplicity, we denote $\widehat{\sigma}^\mathrm{CE}_\mathrm{HH}(f)$ as $\widehat{\sigma}_n(f)$ throughout the remainder of this paper. We observe that when $z^\star=0.08$, the mean subtraction residual begins to get comparable to $\widehat{\sigma}_n(f)$ in the most sensitive band and quickly exceeds it when $z^\star$ keeps increasing. This indicates that, in this band, the uncertainty of the search is no longer dominated by the detector noise, but by the subtraction residual. 

In the right panel of Fig.~\ref{fig: subtraction_vs_std}, we show the subtraction residual of 200 realizations for $z^\star=0.04$ case, where each gray curve corresponds to a particular realization. The $y$-axis shows the relative deviation of the subtraction residual of each realization from the mean subtraction residual. Unsurprisingly, we observe that the subtraction residual is indeed dominated by the “shot-noise” just like the original foreground itself, which makes its statistical description challenging. The loudest subtraction residual is around 14 times louder than the mean subtraction residual foreground. This phenomenon can be intuitively understood: when the original signal is very loud, one can obtain smaller fractional deviations of the estimated parameters from their true values; however, even such small fractional error in parameters may produce a large residual for a loud signal. 

We make two additional remarks here. First, $\sigma_{\Omega_\mathrm{BNS}^{z^\star}}(f)$ will keep decreasing as the observation time $T_\mathrm{obs}$ increases, while the subtraction residual remains roughly constant regardless of how large $T_\mathrm{obs}$ is. Second, Ref.~\cite{Song:2024pnk} demonstrates that including spin parameters in the analysis leads to even larger subtraction residuals. Therefore, the subtraction residual we show here should be considered as an optimistic estimate. 

\begin{figure*}[!htbp]
    \centering
    \includegraphics[width=0.85\linewidth]{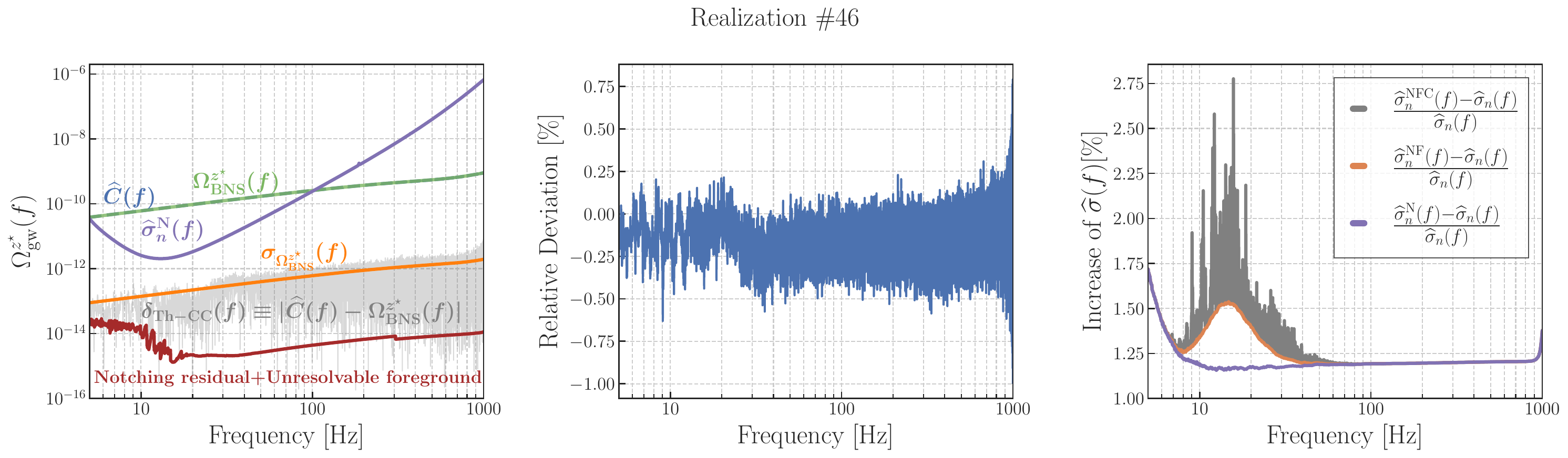}
    \includegraphics[width=0.85\linewidth]{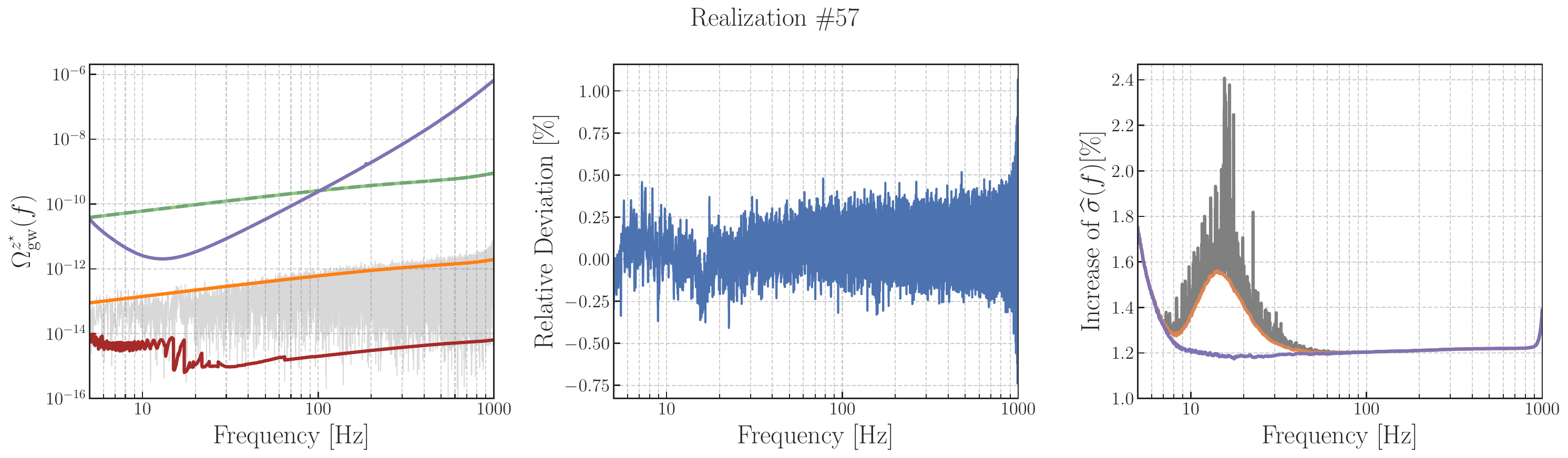}
    \includegraphics[width=0.85\linewidth]{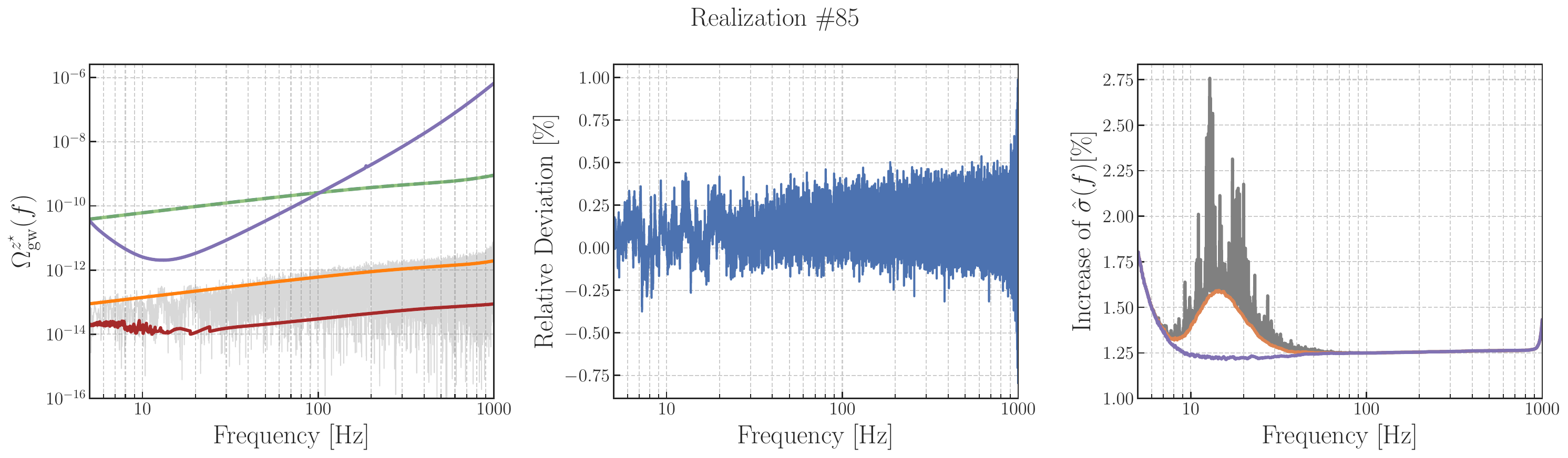}
    \includegraphics[width=0.85\linewidth]{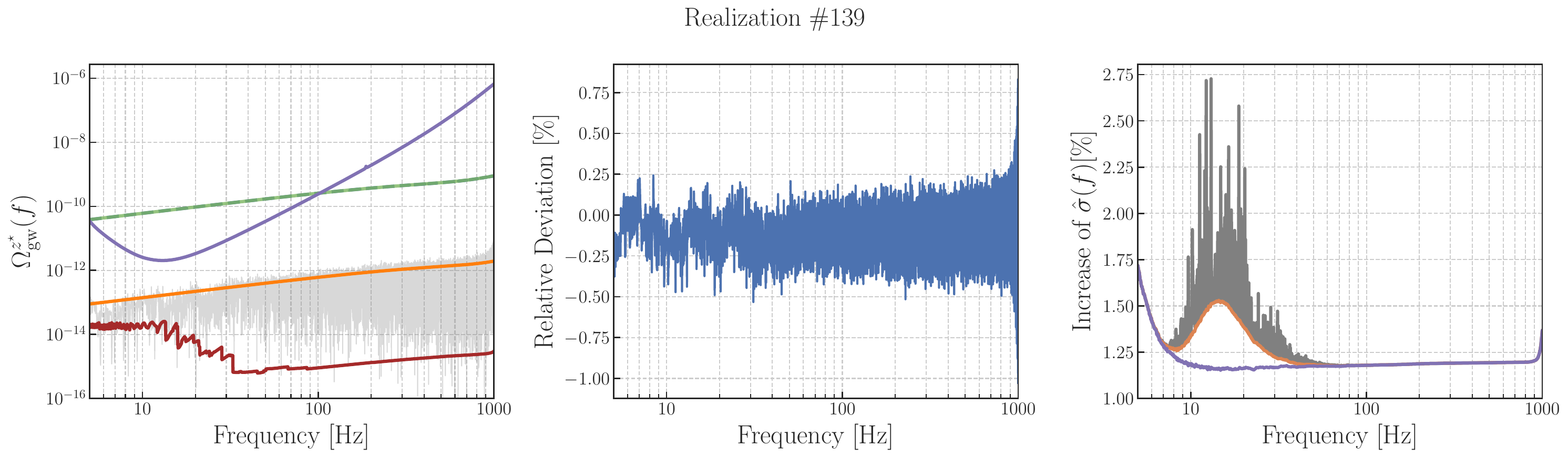}
    \includegraphics[width=0.85\linewidth]{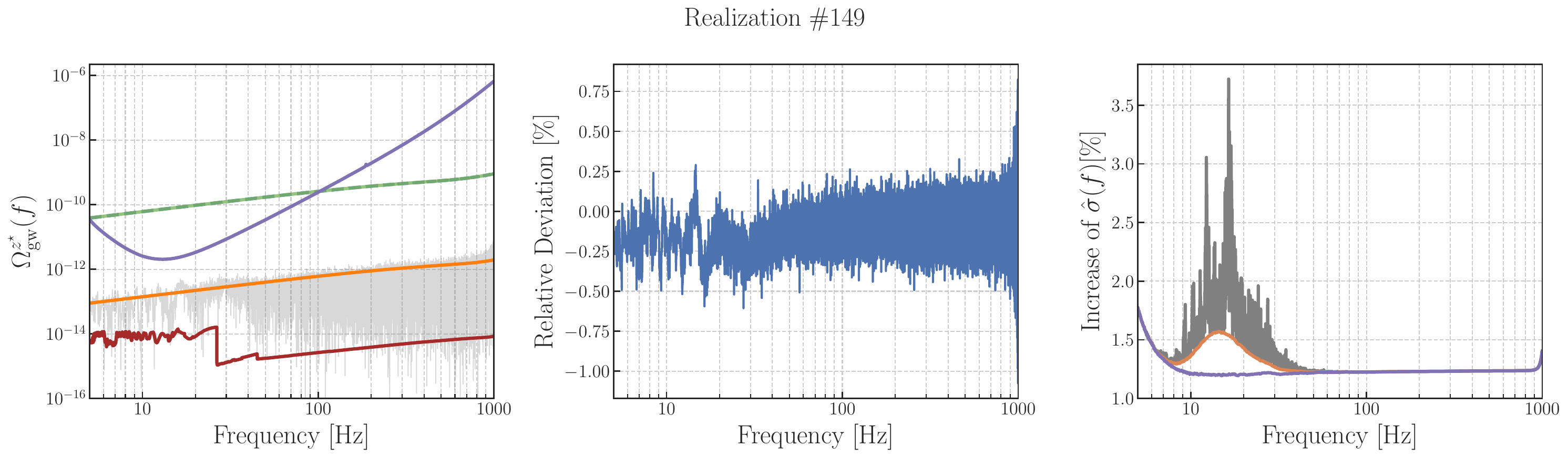}
    \caption{Left: Cross-correlation spectrum $\widehat{C}(f)$ in dashed blue, theoretical estimate of the \ac{BNS} foreground $\Omega_\mathrm{BNS}^{z^\star}(f)$ in green, and the cross-correlation uncertainty after notching, $\sigma_n^\mathrm{N}(f)$ in purple. The gray curves show the absolute difference between $\widehat{C}(f)$ and $\Omega_\mathrm{BNS}^{z^\star}(f)$, and the brown curves represent the notching residual plus the unresolvable foreground. The statistical fluctuation of the foreground is shown in orange. Middle: Relative deviation between $\widehat{C}(f)$ and $\Omega_\mathrm{BNS}^{z^\star}(f)$. Right: Impact on the total uncertainty when incorporating both the statistical fluctuation of the foreground and the deviation between cross-correlation and the theoretical estimate. We fix $z^\star=0.35$ in this analysis.}
    \label{fig: CC}
\end{figure*}

Given the above discussion, we do not adopt the subtraction method for the removal of nearby events. Instead, we turn to the notching method proposed in our previous works~\cite{Zhong:2022ylh,Zhong:2024dss,Zhong:2025qno}, which is also briefly outlined in Appendix~\ref{app: Notching}. As we show in the Ref.~\cite{Zhong:2024dss}, the notching procedure is primarily sensitive to the recovered detector-frame chirp mass--a quantity that can be tightly constrained. Thus, it remains reasonable to neglect spin parameters in this work to simplify the simulations. We also explicitly show the notching residual in Fig.~\ref{fig: Budget} and Fig.~\ref{fig: CC} to prove that one can safely ignore it.

In Fig.~\ref{fig: Budget} we compare the magnitudes of five key quantities: the mean of $\Omega_{\mathrm{BNS}}^{z^\star}(f)$ (blue), $\sigma_{\Omega_\mathrm{BNS}^{z^\star}}(f)$ (orange), the estimated notching residual plus the unresolvable foreground (brown), the unresolvable foreground (green), and the uncertainty of the cross-correlation spectrum $\widehat{C}(f)$ for a one-year observation denoted by $\widehat{\sigma}_n(f)$  (purple). 

As in Fig.~\ref{fig: subtraction_vs_std}, the arrows in Fig.~\ref{fig: Budget} indicate the trends of how corresponding quantities change with increasing $z^\star$. We notice that the unresolvable \ac{BNS} foreground increases with $z^\star$, while the mean and the standard deviation of $\Omega_\mathrm{BNS}^{z^\star}(f)$ decrease. In the above figure, we particularly plot the standard deviation of $\Omega_\mathrm{BNS}(f)$ in dash-dotted line corresponding to $z^\star=0$ case. We observe that if we keep all \ac{BNS} events in the data, then the fluctuation of the foreground itself can be even stronger than the detector noise below 50 Hz, hence the uncertainty of the search can be dominated by $\sigma_{\Omega^{z^\star=0}_\mathrm{BNS}}(f)\equiv\sigma_{\Omega_\mathrm{BNS}}(f)$ instead of $\widehat{\sigma}_n(f)$. However, as long as we specify a $z^\star >0$, the corresponding $\sigma_{\Omega_\mathrm{BNS}^{z^\star}}(f)$ drops down by at least one order of magnitude and stays well below $\widehat{\sigma}_n(f)$.
In addition, the unresolvable \ac{BNS} foreground is roughly two orders of magnitude below the standard deviation of $\Omega_\mathrm{BNS}^{z^\star}(f)$, indicating that it can be safely neglected. As a sanity check, we also plot the estimated notching residual plus the unresolvable \ac{BNS} foreground in brown for the special case $z^\star=0.35$. Because to perform notching procedure for all two hundred realizations is infeasible from the calculation aspect, we randomly draw five realizations and compute the mean of the notching residual plus the unresolvable foreground for them as the brown curve shown in Fig.~\ref{fig: Budget}. Please see Sec.~\ref{sec: CC} for more details and see the following paragraph for the rationale of choosing $z^\star=0.35$.

\begin{figure}
    \centering
    \includegraphics[width=1.0\linewidth]{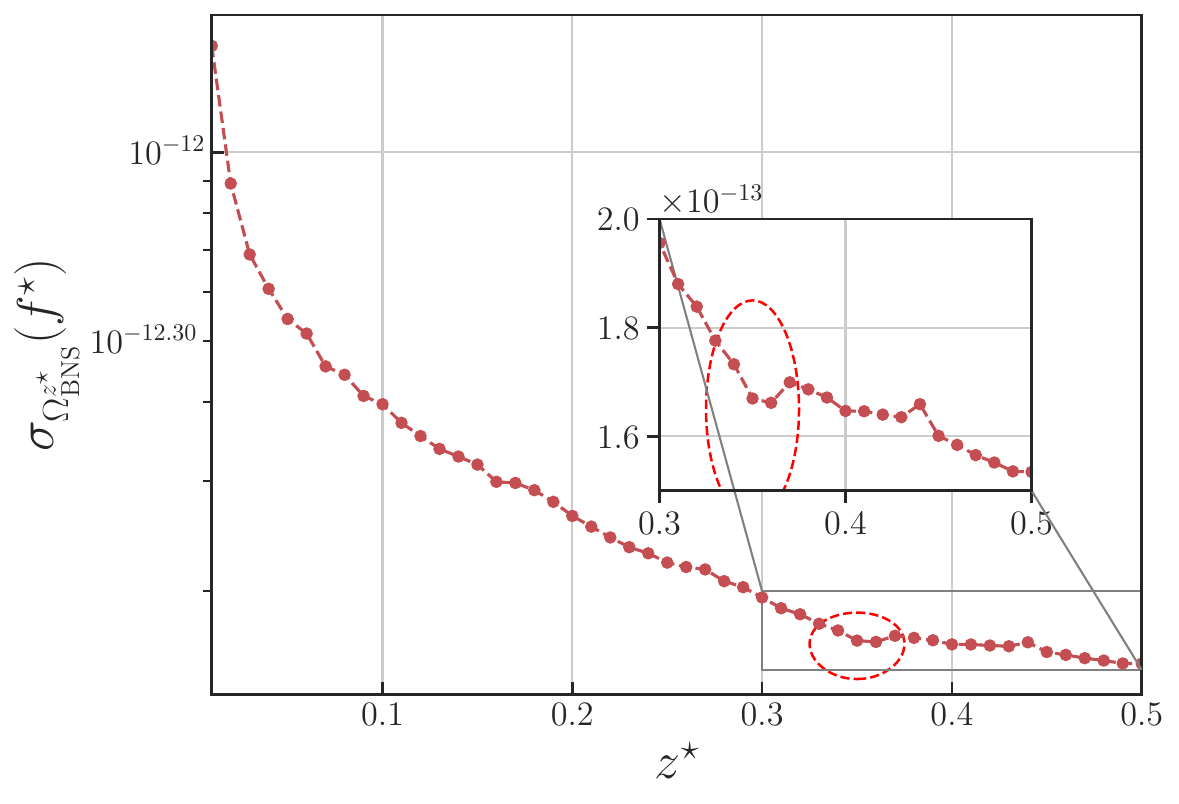}
    \caption{The change of $\sigma_{\Omega_\mathrm{BNS}^{z^\star}}(f)$ at $f=f^\star=13$ Hz as a function of $z^\star$. The standard deviation decreases rapidly for small $z^\star$, then gradually levels off as $z^\star$ approaches about 0.35. The inset and the dashed oval highlight this transition, showing that the rate of decrease becomes minimal beyond this point.}
    \label{fig: sigma_f13}
\end{figure}

To facilitate the subsequent discussion, we examine $\sigma_{\Omega_\mathrm{BNS}^{z^\star}}(f)$ at $f^\star=13$Hz, at which frequency $\widehat{\sigma}_{n}(f)$ reaches its minimum value. We show the behavior of $\sigma_{\Omega_\mathrm{BNS}^{z^\star}}(f^\star)$ in Fig.~\ref{fig: sigma_f13}. Initially $\sigma_{\Omega_\mathrm{BNS}^{z^\star}}(f^\star)$ decreases rapidly as $z^\star$ increases, but the rate of decrease slows significantly as $z^\star$ approaches approximately $0.35$. Based on this trend, we fix $z^\star=0.35$ for the remainder of our analysis. This choice realizes a balance between effectively mitigating the standard deviation of the foreground by removing nearby events and avoiding excessive removal that could lead to a substantial residual background from the notching procedure. 

\section{Bridging the Summation and Cross-Correlation}\label{sec: CC}
In this section, we perform cross-correlation analyses to explicitly estimate the energy density of the \ac{BNS} foreground $\widehat{C}(f)$ and compare it with the theoretical summation result $\Omega_\mathrm{BNS}^{z^\star}(f)$ computed by Eq.~\eqref{eq: summation_zstar}. 

Due to the extreme computational cost of injecting all 200 realizations of approximately $\sim10^8$ \ac{BNS} events into time series and performing cross-correlation, we randomly select five realizations for this test. For each selected realization, we consider events with redshift greater than $z^\star=0.35$ and inject them into timeseries using the \textsc{IMRPhenomXAS}~\cite{IMRXAS} waveform model by \textsc{Bilby}~\cite{Bilby} without injecting noise series. We then apply \textsc{Pygwb}~\cite{pygwb} to perform cross-correlation analysis. In this test, we choose the segment length $T=192$ s, the frequency resolution $\delta f=1/32$ Hz, and adopt 50\% overlapping segments with Hann windowing, which is the standard choice for the stochastic search~\cite{O3stoch, pygwb}. The results are shown in Fig.~\ref{fig: CC}. 

The first column shows the cross-correlation results in dashed blue, the theoretical estimation of $\Omega_\mathrm{BNS}^{z^\star}(f)$ in green, and the difference between the two in gray. To simplify our notation, we denote the absolute value of this difference by $\delta_\mathrm{Th-CC}(f)$.  In addition, we plot $\sigma_{\Omega^{z^\star}_\mathrm{BNS}}(f)$ in orange, and show the notching residual plus the unresolvable foreground in brown. For comparison, we also show the predicted $\widehat{\sigma}_n^\mathrm{N}(f)$ in purple, representing the uncertainty of $\widehat{C}(f)$ after notching out nearby resolvable events.

To obtain these brown curves and $\widehat{\sigma}_n^\mathrm{N}(f)$, we inject \textit{all} of the events with redshift less than $z^\star=0.35$ into the time series and then perform the cross-correlation computation. We then use the procedures detailed in Appendix~\ref{app: Fisher} and Appendix~\ref{app: Notching} to draw recovered parameters for the resolvable events among these nearby events, and determine corresponding masks. These masks are applied to time-frequency spectrograms, and the remaining pixels are combined to compute the notching residual plus the unresolvable foreground and its corresponding uncertainty $\widehat{\sigma}^\mathrm{N}_n(f)$ based on Eq.~\eqref{eq:combine_t}. We note that although the data we use for cross-correlation is \textit{noise-free}, we can still estimate $\widehat{\sigma}_n^\mathrm{N}(f)$ by Eq.~(\ref{eq:stoch}-\ref{eq:combine_t}) given detector noise \ac{PSD}. As a reminder, removing time-frequency pixels leads to a loss of information, which in turn increases $\widehat{\sigma}(f)$, i.e. $\widehat{\sigma}_n^\mathrm{N}(f)\geqslant \widehat{\sigma}_n(f)$.

From this set of plots, we conclude that the notching residual plus the unresolvable foreground is indeed smaller than $\sigma_{\Omega_\mathrm{BNS}^{z^\star}}(f)$ by at least an order of magnitude and can thus be neglected. On the other hand, the discrepancy between $\widehat{C}(f)$ and $\Omega_\mathrm{BNS}^{z^\star}(f)$ is comparable to $\sigma_{\Omega_\mathrm{BNS}^{z^\star}}(f)$ and should therefore be accounted for. 


In the second column, we show the relative deviation of the cross-correlation result from the theoretical estimation, expressed as a percentage. Across all five realizations, we observe that this relative deviation remains below approximately $0.5\%$ over the entire frequency range. Moreover, we aim to estimate how taking $\sigma_{\Omega_\mathrm{BNS}^{z^\star}}(f)$ and $\delta_\mathrm{Th-CC}(f)$ into account can reduce the sensitivity of the search. Before proceeding, we define $\widehat{\sigma}_n^\mathrm{NF}(f)$ and $\widehat{\sigma}_n^\mathrm{NFC}(f)$ as following
\begin{equation}
    \widehat{\sigma}_n^\mathrm{NF}(f):=\sqrt{(\widehat{\sigma}_n^\mathrm{N})^2(f)+\sigma^2_{\Omega_\mathrm{BNS}^{z^\star}}(f)},
    \label{eq: sigma^NF}
\end{equation}
\begin{equation}
    \widehat{\sigma}_n^\mathrm{NFC}(f):=\sqrt{(\widehat{\sigma}_n^\mathrm{N})^2(f)+\sigma^2_{\Omega_\mathrm{BNS}^{z^\star}}(f)+\delta^2_\mathrm{Th-CC}(f)}.
    \label{eq: sigma^NFC}
\end{equation}
We emphasize that $\sigma_n^\mathrm{NFC}(f)$ serves as an \textit{ad hoc} estimate of the overall sensitivity of the search, incorporating the various new sources of uncertainties discussed above. 

In the third column of the plot, we show the relative increase of $\widehat{\sigma}(f)$ after accounting for the effects of notching, the statistical fluctuation of the foreground, and the deviation between the cross-correlation result and the theoretical estimate of $\Omega_\mathrm{BNS}^{z^\star}(f)$. We observe that notching out resolvable events increases $\widehat{\sigma}(f)$ by approximately $\gtrsim1.25\%$ across the entire frequency band. Because we remove most of pixels in the low frequency range around $f\sim 5$ Hz, the purple curves exhibit a peak below 10 Hz. Due to the exact same reason, we observe the first peak at 5 Hz for orange curves. The second peak of orange curves, near 15 Hz, arises from the statistical fluctuation in the foreground ($\sigma_{\Omega^{z^\star}_{\mathrm{BNS}}}(f)$) and the fact that the frequency range $\sim10-30$ Hz corresponds to the most sensitive band for the detectors configuration considered in this study. Based on the same reason, the deviation between the cross-correlation and the theoretical estimate ($\delta_{\mathrm{Th-CC}}(f)$) has the largest impact on $\widehat{\sigma}_n^\mathrm{NFC}(f)$ in this frequency band as shown by gray curves. 

For all five realizations, the combined effect of the factors considered leads to an increase in $\widehat{\sigma}_n(f)$ of at most $\lesssim 4\%$ (caused by notching, $\sigma_{\Omega_\mathrm{BNS}^{z^\star}}(f)$, and $\delta_{\mathrm{Th-CC}}(f)$) below 40 Hz, and about $\sim 1\%$ (dominated by notching only) above 40 Hz. We stress that in a real search, one has access to $\widehat{\sigma}_n^\mathrm{N}(f)$ after notching out resolvable events, hence to account for other factors in estimating $\widehat{\sigma}^\mathrm{NCF}_n(f)$, one can multiply $\widehat{\sigma}_n^\mathrm{N}(f)$ by a function $F(f)$
\begin{equation}
    F(f):=
    \left\{
    \begin{aligned}
        &1.03,~f\leqslant 40~\mathrm{Hz}\\
        &1.00,~f> 40~\mathrm{Hz}.
    \end{aligned}
    \right.
\end{equation}

Before closing this section, we leave two side remarks here: 
\begin{enumerate}
    \item[(1)] $\widehat{\sigma}_n(f)$ in the above calculations assumes two co-located and co-aligned detectors. If instead the detectors were placed separately at LIGO Hanford site and LIGO Livingston site, $\widehat{\sigma}_n(f)$ would be larger across the entire band. Consequently, the increase in $\widehat{\sigma}(f)$ shown in the third column of Fig.~\ref{fig: CC} would in turn be smaller. In other words, the sensitivity reduction we report is conservative and likely represents an upper bound.
    \item[(2)] We emphasize that in the simulations presented above, we assume knowledge of the true noise \ac{PSD}. In practice, however, the true \ac{PSD} is not known \textit{a priori} and must be estimated from the data. In Appendix~\ref{app: weighted_avg}, we test the impact of using an estimated \ac{PSD} for the cross-correlation analysis. We find that using the estimated \ac{PSD} leads to an overestimation of the search uncertainty by approximately $\lesssim 4\%$ in the 10–40 Hz frequency band. On the other hand, as discussed earlier, the search uncertainty may also be underestimated due to neglecting other factors such as statistical fluctuation of the foreground and the discrepancy between $\widehat{C}(f)$ and $\Omega_\mathrm{BNS}^{z^\star}(f)$. These opposing effects can partially cancel each other out. For further details, we refer the reader to Appendix~\ref{app: weighted_avg}.
\end{enumerate}

\section{Discussions and Conclusions}\label{sec: Dis_Conc}
In this work, we discuss the impact of nearby \ac{BNS} events, also known as shot noise, to the overall \ac{BNS} foreground. By simulating two hundred realistic one-year long realizations of \ac{BNS} events, we observe that the shot noise can dominate $\Omega_\mathrm{BNS}(f)$. In other words, $\Omega_\mathrm{BNS}(f)$ for different realizations can have very different amplitudes. Hence, one should not simply take the average of $\Omega_\mathrm{BNS}(f)$ from many realizations as the model prediction of the corresponding magnitude of the \ac{BNS} foreground for the downstream Bayesian analysis.

To mitigate the impact caused by shot noise, we define $\Omega_\mathrm{BNS}^{z^\star}(f)$, which is the energy density of the \ac{BNS} foreground after removing events whose redshift is less than $z^\star$. As presented by Fig.~\ref{fig: Deviation_BNS_z_star} and Fig.~\ref{fig: Deviation_BNS_z_star_3freqs}, $\Omega_\mathrm{BNS}^{z^\star}(f)$ is much better behaved than the original $\Omega_\mathrm{BNS}(f)$, and the statistical fluctuation in $\Omega_{\mathrm{BNS}}^{z^\star}(f)$ can be approximately characterized by a simple Gaussian. We then studied two approaches to remove these neighboring events in practice. We investigated the possibility of subtracting these events out in the frequency domain: as shown by Fig.~\ref{fig: subtraction_vs_std} and Fig.~\ref{fig: Budget}, the subtraction residual increases very fast as $z^\star$ increases, and becomes greater than the $\sigma_{\Omega_\mathrm{BNS}^{z^\star}}(f)$ when $z^\star=0.04$. There are two other issues we recognized associated with the subtraction residual: (1) As shown by Fig.~\ref{fig: subtraction_vs_std}, the residual itself is still dominated by shot noise, hence making describing the residual statistically challenging; (2) As pointed out in Ref.~\cite{Song:2024pnk}, when considering spin parameters, the resulting subtraction residual can be even larger, which means that the residual foreground we computed here is an optimistic estimate. As a result, we turn to an alternative method to remove nearby events---notching method proposed in our previous work~\cite{Zhong:2022ylh, Zhong:2024dss,Zhong:2025qno}, which only relies on a precise estimation of the detector-frame chirp mass of the resolvable \ac{CBC} systems~\cite{Zhong:2024dss}.

In Sec.~\ref{sec: CC}, we further consider the possible deviation of the cross-correlation estimator from the theoretical estimation of $\Omega_\mathrm{BNS}^{z^\star}(f)$. We randomly picked five realizations among two hundred realizations, injected those \ac{BNS} events into the time series, and performed cross-correlation calculation to get $\widehat{C}(f)$ as shown in Fig.~\ref{fig: CC}. Fig.~\ref{fig: CC} shows the good agreement between the theory and the practical measurement, and also directly shows that the notching residual and the unresolvable foreground are indeed negligible (see brown curves). To investigate the impact of notching, statistical fluctuation of the foreground itself and the deviation of the cross-correlation spectrum from the theoretical estimation on the overall sensitivity of a stochastic search, we defined $\sigma_n^\mathrm{NF}(f)$ and $\sigma_n^\mathrm{NFC}(f)$ as Eq.~\eqref{eq: sigma^NF} and Eq.~\eqref{eq: sigma^NFC}
. We find that the resulting sensitivity loss in the search for an isotropic background formed by binary neutron star mergers is minimal, and is limited to $\lesssim 4\%$ below 40 Hz, and to $\lesssim 1\%$ above 40 Hz. 

In conclusion, we claim that $\Omega_\mathrm{BNS}^{z^\star}(f)$ is the right quantity to use for the subsequent Bayesian analysis to estimate population hyperparameters $\bm\Lambda$, and the corresponding likelihood, which is intensively used in the literature and given by Eq.~\eqref{eq: likelihood_old} should be modified as following 
\begin{equation}
\begin{aligned}
     \mathscr{L}&(\widehat{C}^\mathrm{N}(f)|\bm{\Lambda})=\\
     &\frac{1}{\sqrt{2\pi}\widehat{\sigma}_n^{\mathrm{NFC}}(f)} \exp\left(-\frac{(\widehat{C}^\mathrm{N}(f)-\langle\Omega_\mathrm{BNS}^{z^\star}(f|\bm\Lambda)\rangle)^2}{2(\widehat{\sigma}_n^\mathrm{NFC})^2(f)}\right),
\end{aligned}
\end{equation}
where $\widehat{C}^\mathrm{N}(f)$ denotes the cross-correlation spectrum after notching out nearby resolvable \ac{BNS} events, and the corresponding uncertainty $\widehat{\sigma}_n^\mathrm{NFC}(f)$ is defined as Eq.~\eqref{eq: sigma^NFC}. $\langle\Omega_\mathrm{BNS}^{z^\star}(f|\bm\Lambda)\rangle)$ denotes the usual ensemble average of the foreground energy density, which can be computed using Eq.~\eqref{eq: summation_zstar} by simulating a large number of \ac{BNS} events with redshift $z\geqslant z^\star$.

We highlight that although we focus on \ac{BNS} system only in this work, and only consider one particular population, the methodology we proposed can be easily generalized to any \ac{CBC} systems with different populations, and also can be directly adopted to any astrophysical \ac{GW} sources with finite rate.
\section*{Acknowledgments}
The authors are grateful for Joseph D. Romano for reading the first draft of this manuscript and providing precious comments and suggestions. The authors are grateful for computational resources provided by the LIGO Laboratory and supported by National Science Foundation (NSF) Grants PHY-0757058 and PHY-0823459. The authors are grateful for computational resources provided by Cardiff University, and funded by an STFC grant supporting UK Involvement in the Operation of Advanced
LIGO.

\appendix
\renewcommand{\theequation}{\thesection.\arabic{equation}}
\counterwithin{equation}{section}
\begin{figure*}[!htbp]
    \centering
    \includegraphics[width=1.0\linewidth]{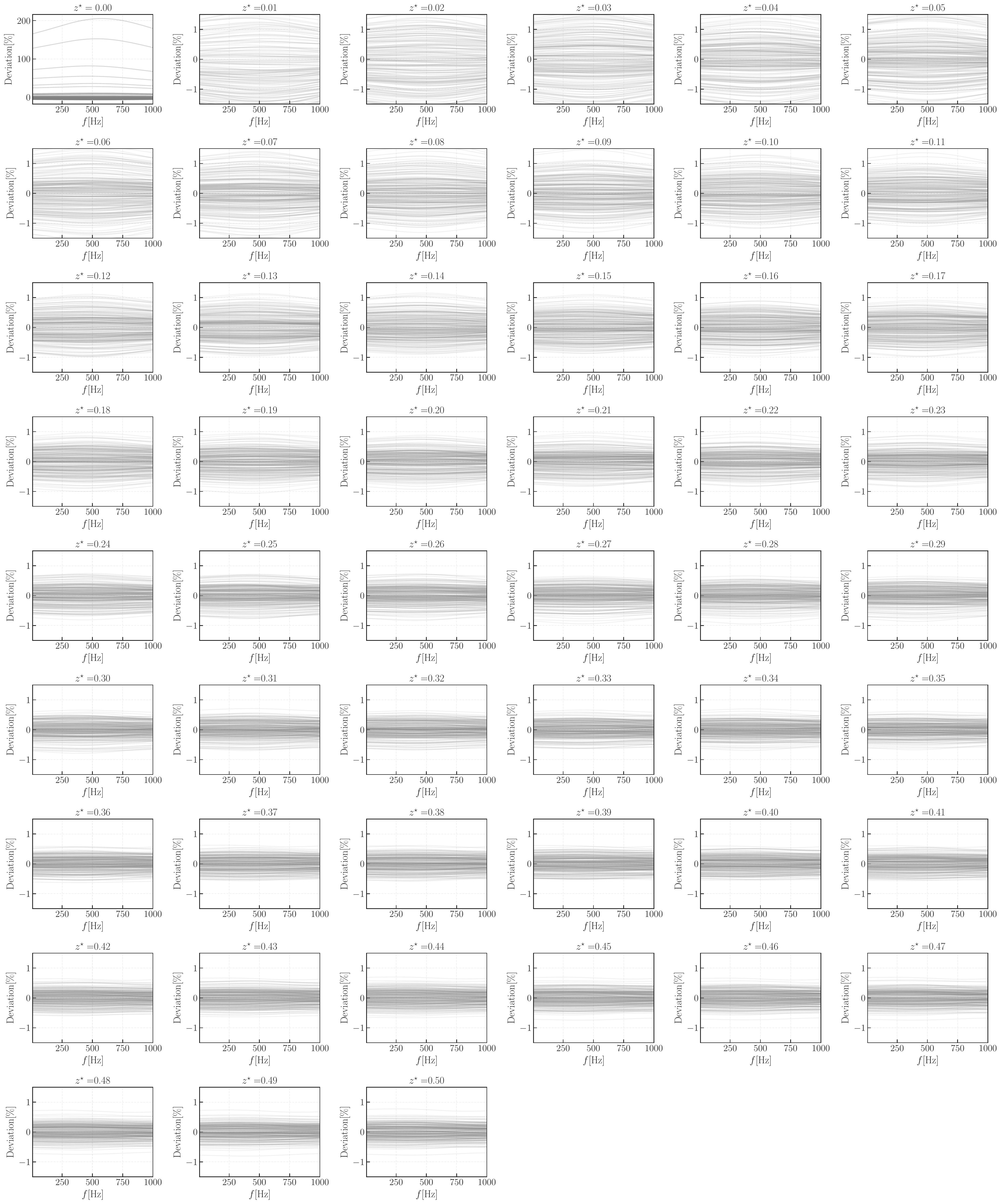}
    \caption{Complete version of Fig.~\ref{fig: Deviation_BNS_z_star_six}}
    \label{fig: Deviation_BNS_z_star}
\end{figure*}
\section{Extra Plots}\label{app: extra_plots}
\begin{figure*}[!htbp]
    \centering
    \includegraphics[width=1.0\linewidth]{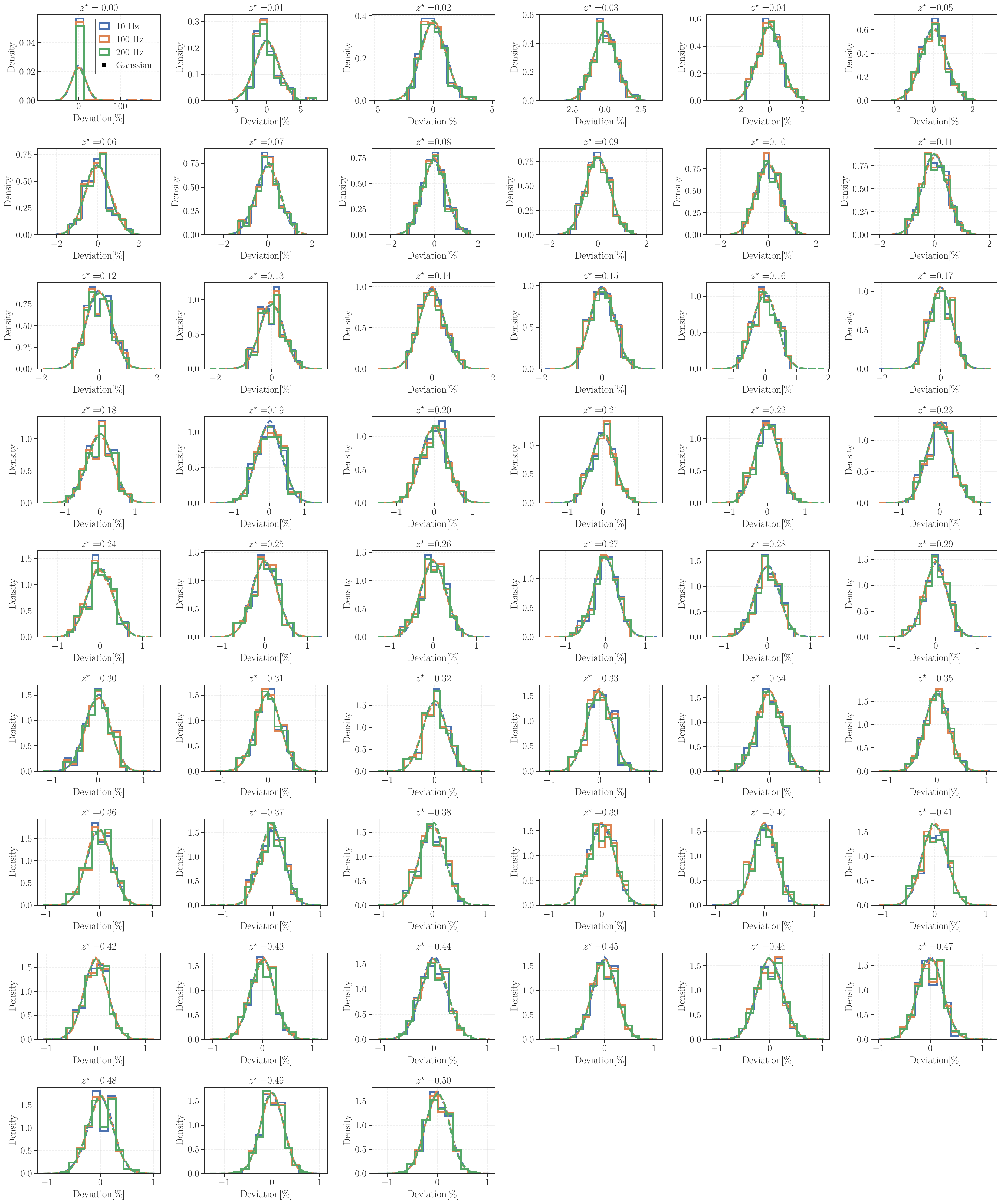}
    \caption{Complete version of Fig.~\ref{fig: Deviation_BNS_z_star_3freqs_six}}
    \label{fig: Deviation_BNS_z_star_3freqs}
\end{figure*}

We show two extra plots (Fig.~\ref{fig: Deviation_BNS_z_star} and Fig.~\ref{fig: Deviation_BNS_z_star_3freqs}) in this appendix.
\section{Fisher Analysis}\label{app: Fisher}
In this appendix, we briefly introduce the Fisher analysis, please see Ref.~\cite{Zhou:2022nmt, Zhou:2022otw} for more details. We consider a signal $h$ to be individually resolved if its network \ac{SNR}
\begin{equation}
    \rho = \sqrt{\sum_{J=1}^{N_\mathrm{det}} \langle h|h\rangle_J}
\end{equation}
is above a certain threshold, where $\langle\cdot|\cdot\rangle$ denotes the usual signal inner product
\begin{equation}
\langle a|b\rangle = 4\,\mathrm{Re} \int_0^{\infty}\frac{a(f) b^*(f)}{P_n(f)}\,\D f \,,
\label{eq_fisher}
\end{equation}
and the sum index $J$ runs over all the $N_\mathrm{det}$ detectors in the network. $P_n(f)$ denotes the detector noise \ac{PSD}, which is assumed to be \textit{known} in the main text of this work. Same as our previous work Ref.~\cite{Zhong:2022ylh, Zhong:2025qno}, we choose $\rho_\mathrm{th}=12$. We estimate the impact of the imperfect recovery of the signals on the notching procedure via their Fisher matrix~\cite{Finn:1992wt}
\begin{equation}
\Gamma_{\alpha\beta} = \sum_{J=1}^{N_\mathrm{det}}\left\langle\frac{\partial h}{\partial\theta^\alpha} \middle| \frac{\partial h}{\partial\theta^\beta}\right\rangle_J \,,
\label{eq:fisher}
\end{equation}
with $\theta^\alpha$ denoting the source parameters. We consider a set of 9 parameters ignoring spin of all \acp{CBC}
\begin{equation}
\bm{\theta}=
\left\{ 
m_1^d, 
m_2^d, 
d_L, 
\iota, 
\delta,
\alpha, 
\psi,
\phi_{c}, 
t_{c} 
\right\} \,.
\label{eq:info_paras}
\end{equation}
Here, $m_1^d = m_1 (1+z)$, $m_2^d=m_2(1+z)$ are the detector-frame primary and secondary masses; $d_{L}$ is the luminosity distance; $\iota,\psi$ are the inclination and polarization angles; $\alpha,\delta$ are the right ascension and declination angles; $\phi_c, t_c$ are the phase and time of coalescence. We compute the Fisher matrices with the public package \texttt{GWFish}~\cite{DupletsaHarms2023}.

For each resolvable \ac{BNS} event ($\rho>12$), we sample its source parameters \textit{once} from a multivariate normal distribution $\mathcal{N}(\widehat{\bm{\theta}},\Gamma^{-1})$, with mean located at the true values of the parameters $\widehat{\bm{\theta}}$. We note that this procedure is not intended to produce \textit{a single} posterior sample of source parameters, but rather a single \textit{measurement}.

\section{Cross-correlation}\label{app: CC}
We introduce cross-correlation method in this appendix, and we refer to Ref.~\cite{Allen:1997ad, Romano:2016dpx, Zhong:2025} for more details.
\begin{figure*}
    \centering
    \includegraphics[width=1.0\linewidth]{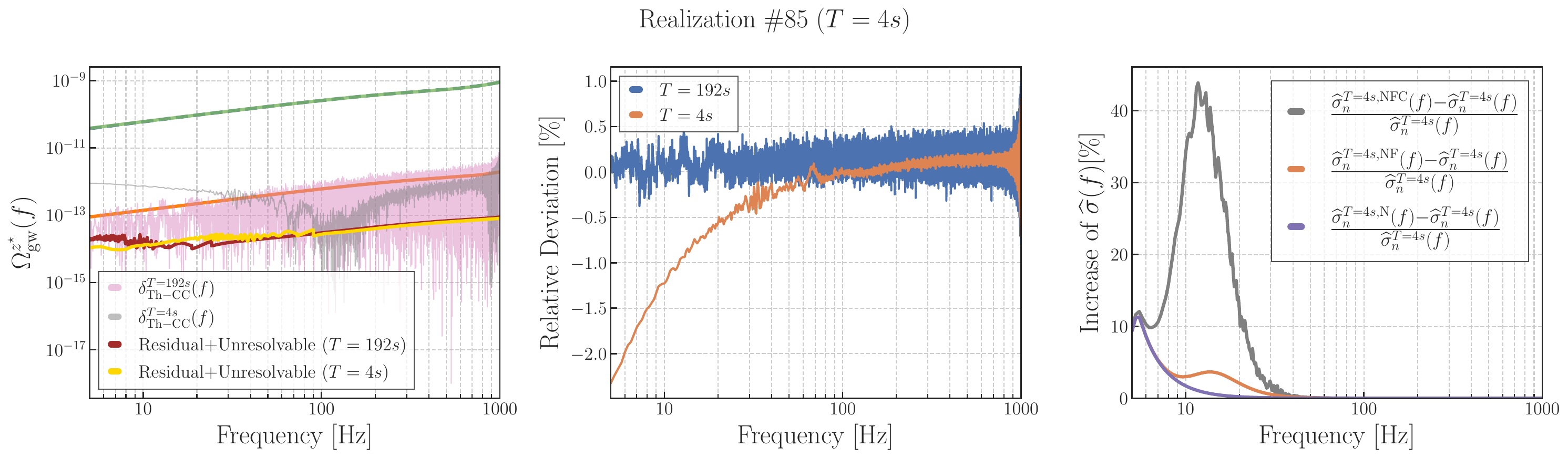}
    \caption{Similar to Fig.~\ref{fig: CC}, but here we consider to use $T=4$ s, $\delta f=1/4$ Hz for cross-correlation calculation.}
    \label{fig: 4scase}
\end{figure*}
Let us consider two \ac{GW} detectors $I, J$ with strain time series $h_I(t)$, $h_J(t)$. To perform the cross-correlation search for the \ac{SGWB}, we split the strains into time segments $t_i$ of duration $T=192\,\rm{s}$. For each frequency bin $f_j$, we denote the Fourier transforms of these segments by $\tilde{h}_{I,J}(t_i;f_j)$ and their complex conjugates with an asterisk. For the baseline of these two detectors, one can define the cross-correlation statistics $\widehat{C}_{IJ}(t_i;f_j)$ and the corresponding variance estimator $\widehat{\sigma}_{IJ}^2$ as~\cite{KAGRA:2021kbb, Allen:1997ad}
\begin{eqnarray}
    \widehat{C}_{IJ}(t_i;f_j) & = & \Big( \frac{20\pi^2f_j^3}{3H_0^2T} \Big) \frac{\Re[{\tilde{h}_{\rm{I}}^*(t_i;f_j)\tilde{h}_{\rm{J}}(t_i;f_j)}]} {\gamma_{IJ}(f_j)}\,, \nonumber \\
    \widehat{\sigma}_{IJ}^2(t_i;f_j) & = & \Big( \frac{20\pi^2f_j^3}{3H_0^2} \Big)^2 \frac{P_{n_{\rm{I}}}(t_i;f_j)P_{n_{\rm{J}}}(t_i;f_j)} {8T\delta f\gamma_{{IJ}}^2(f_j)}\,.
    \label{eq:stoch}
\end{eqnarray}
Here $\gamma_{{IJ}}(f_j)$ is the overlap reduction function between the two detectors~\cite{Allen:1997ad}, $\delta f=1/32$\,Hz is the Fourier transform resolution, $P_{nI}(t_i;f_j)$ is the \ac{PSD} of detector $I$ at time $t_i$, and $H_0$ is the Hubble constant. The time-averaged frequency-domain spectrum  $\widehat{C}_{IJ}(f_j)$ and its variance $\widehat{\sigma}_{IJ}^2(f_j)$ can then be obtained by performing a weighted average over all time segments:
\begin{eqnarray}
\widehat{C}_{IJ}(f_j) & = & \frac{\displaystyle{\sum_{i} \widehat{C}_{IJ}(t_i;f_j) \widehat{\sigma}_{IJ}^{-2}(t_i;f_j)}} {\displaystyle{\sum_{i} \widehat{\sigma}_{IJ}^{-2}(t_i;f_j)}}\,,\nonumber\\
\widehat{\sigma}_{IJ}^{-2}(f_j) & = & \sum_{i} \frac{1}{\widehat{\sigma}^2_{IJ}(t_i;f_j)}\,.
\label{eq:combine_t}
\end{eqnarray}
With a network of more than two detectors, these spectra can be further combined by computing weighted averages among all the possible detector pairs. We refer Ref.~\cite{Zhong:2025} for more technical details about the statistical properties of $\widehat{C}_{IJ}(t_i,f_j)$.

\section{Notching method}\label{app: Notching}
In this appendix, we introduce the notching procedure in a nutshell, and we refer Ref.~\cite{Zhong:2022ylh, Zhong:2025} for more technical details.

After drawing recovered samples for all resolvable \ac{BNS} events, one can inject these events into time series and perform cross-correlation calculation. Note that, since we do not inject noise into these time series, one should be able to nail down the energy distribution of all of these recovered signals in the time-frequency domain straightforwardly. Different from how we define masks in our previous paper Ref.~\cite{Zhong:2024dss, Zhong:2025qno}, we adopt a slightly different way to define masks here. In this work, we directly set a threshold for $|C(t_i;f_j)|$ spectrograms, where as long as the pixel value is higher than the threshold, then the corresponding pixel is labeled to be notched out. In this work, the threshold is chosen as $10^{-13}$. We note that this is a tunable parameter: one could choose a smaller threshold to mitigate the notching residual, but also could remove more pixels and hence lose sensitivity.

\section{4s analysis v.s. 192s analysis}\label{app: 4svs192s}
In our previous works (c.f. Ref.~\cite{Zhong:2022ylh, Zhong:2024dss, Zhong:2025qno}) we chose $T=4$s and $\delta f=1/4$Hz for the cross-correlation calculation, but we turn to use $T=192$s and $\delta f=1/32$ Hz in this work. We compare these two choices in this appendix. 

Similar to Fig.~\ref{fig: CC}, we show the results for the analysis using $T=4$ s and $\delta f=1/4$ Hz in the Fig.~\ref{fig: 4scase}. In the first panel, we again show the comparison between theoretically computed $\Omega_\mathrm{BNS}^{z^\star}(f)$ and cross-correlation result. The absolute difference between two are shown in gray, and we overplot this quantity for the $T=192$s case in pink for reference. We also plot notching residual and the unresolvable foreground in brown and gold for $T=192$ s case and $T=4$ s case respectively, where two curves almost coincide with each other. It is notable that $\delta_\mathrm{Th-CC}^{T=4s}>\delta_\mathrm{Th-CC}^{T=192s}$ when $f\lesssim 40$ Hz, which could be due to two reasons. First, in the low frequency band, \ac{BNS} signals are almost monochromatic, and the frequency evolution of the binary is rather slow. For instance, for a \ac{BNS} system composed of two $1.4M_\odot$ neutron stars at redshift $z=1$, the time for the binary to evolve from $f=5$ Hz to $f=6$ Hz can take more than 10 minutes in the detector frame. Thus, a high frequency resolution of the analysis is highly necessary for us to capture the energy density spectrum of the signal precisely. Otherwise, the power of the signal can leak into the nearest bins and cause a bias. Besides, the cross-correlation search is optimal for a persistent, stationary and Gaussian \ac{SGWB}, while the \ac{BNS} foreground violates these assumptions. Using a short segment length can render the data highly non-stationary due to transients, while a longer segment length better preserves stationarity of the data.

In the middle panel, we present the deviation of the cross-correlation result from the theoretical prediction for both the $T=4$s (orange) and $T=192$s (blue) analyses.  The orange curve deviates from 0 by approximately $\gtrsim 1\%$ for $f\lesssim 10$ Hz and gradually approaches to zero at high frequencies. In conclusion, the $T=4$ s analysis with a lower frequency resolution is affected, while the $T=192$ s analysis with a higher frequency resolution is only minimally impacted.
\begin{figure}[!hbp]
    \centering
    \includegraphics[width=0.8\linewidth]{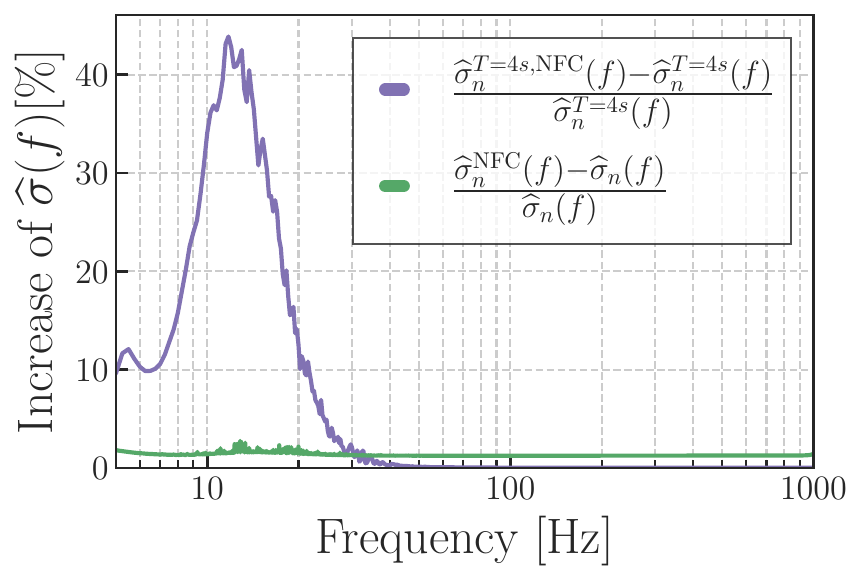}
    \caption{Comparison of the increase of $\widehat{\sigma}(f)$ for $T=4$ s, $\delta f=1/4$ Hz analysis (purple) and $T=192$ s, $\delta f=1/32$ Hz analysis (green).}
    \label{fig: 4svs192s}
\end{figure}
\begin{figure*}[!htbp]
    \centering
    \includegraphics[width=1.0\linewidth]{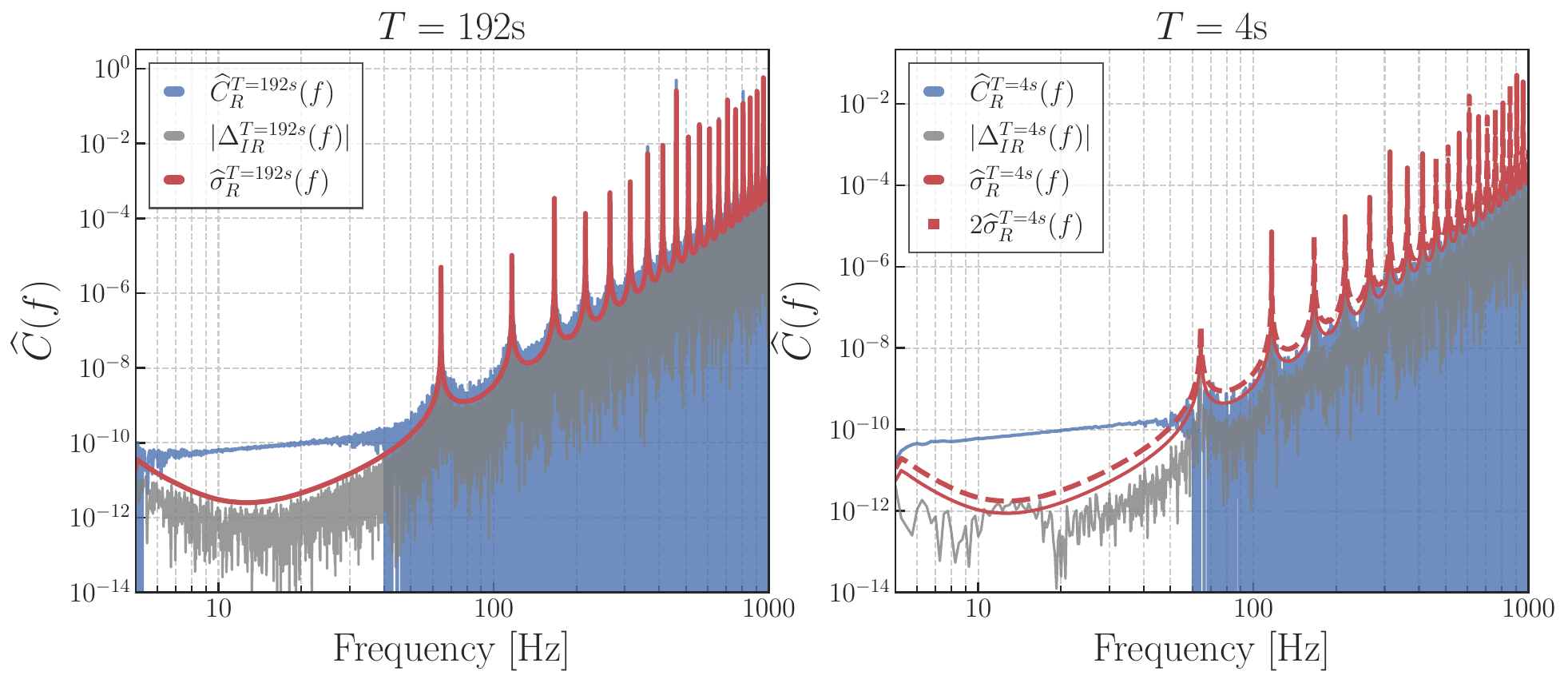}
    \caption{Cross-correlation result for the Realization \# 172, where a detector network consisting of two detectors located at LIGO Hanford and LIGO Livingston with CE2 sensitivity is considered. }
    \label{fig: Real172}
\end{figure*}

Finally, in the right panel, we show how much $\widehat{\sigma}(f)$ increases in the $T=4$ s and $\delta f=1/4$ Hz analysis after all the factors we discussed in the main text are accounted for. Similar to what have seen in Fig.~\ref{fig: CC}, pixels removal is concentrated at low frequencies, whereas in this analysis, fraction of pixels removed with $f\gtrsim 20$ Hz is minimal. More quantitatively, $\widehat{\sigma}_n^{T=4s,\mathrm{N}}(f)$ rises by roughly $10\%$ near $f\simeq 5$ Hz relative to $\widehat{\sigma}_n^{T=4s}(f)$; the corresponding increase for the $T=192$ s analysis is only about $1.8\%$. At higher frequencies, the situation reverses: $\widehat{\sigma}_n^{T=4s,\mathrm{N}}(f)$ is essentially unchanged, whereas $\widehat{\sigma}_n^{T=192s,\mathrm{N}}(f)$ grows by around $1.25\%$ relative to $\widehat{\sigma}_n^{T=192s}(f)$ (see the right-most subplot of the third column of Fig.~\ref{fig: CC}). This difference again arises from the contrasting frequency resolutions and segment lengths of two analyses. Considering the low frequency band where the \ac{BNS} signals can stay for rather long time (can up to several hours), the poor frequency resolution of the analysis with $\delta f=1/4$ Hz causes the signal power to concentrate in only a few bins. Nevertheless, the analysis with $\delta f=1/32$ Hz distributes the signal power over eight times more bins, creating a “dilution” effect: fewer bins are significantly affected ($10\%\simeq 8\times1.8\%$). While at the high frequency range, the analysis with a shorter segment duration is able to localize each \ac{BNS} “track” more precisely in the time-frequency plane, so fewer pixels are notched out, but for the analysis with longer segments, even a few $\mathcal{O}(1)$ s-long transients can lead to removal of an entire 192 s-long pixel.

Now turn to the gray curve in the right-most panel of Fig.~\ref{fig: 4scase}, we notice that $\widehat{\sigma}_n^{T=4s,\mathrm{NFC}}(f)$ increases by $\gtrsim40\%$ in the most sensitive band, whereas in the $T=192$s case, the increase is $\lesssim3\%$. To explain this difference, we first observe that $\delta^{T=4s}_{\mathrm{Th-CC}}(f)>\delta^{T=192s}_{\mathrm{Th-CC}}(f)$ in this frequency band because of two potential reasons we discussed above. In addition, we have $\widehat{\sigma}^{T=4s}_n(f)<\widehat{\sigma}^{T=192s}_n(f)$, which further makes $\delta^{T=4s}_\mathrm{Th-CC}(f)/\widehat{\sigma}^{T=4s}_n(f)>\delta^{T=192s}_\mathrm{Th-CC}(f)/\widehat{\sigma}^{T=192s}_n(f)$, so results in a bigger relative increase of $\widehat{\sigma}(f)$.

We highlight that in the Fig.~\ref{fig: 4scase} and discussions above, we especially denote the uncertainty $\widehat{\sigma}_n(f)$ of the analysis with $T=4$ s and $\delta f=1/4$ Hz by $\widehat{\sigma}_n^{T=4s}(f)$ to indicate that it does not equal to $\widehat{\sigma}_n^{T=192s}(f)$. In fact, $\widehat{\sigma}_n^{T=192s}(f)$ is larger than $\widehat{\sigma}_n^{T=4s}(f)$ across all frequency bins. However, this does not imply that the longer-segment analysis has worse overall sensitivity; it simply reflects that the uncertainty in each frequency bin is larger due to the finer frequency resolution. In a Bayesian framework, the width of the posterior distribution for the parameters of interest  should remain consistent regardless of $T$ and $\delta f$, as the underlying time-domain data data $h(t)$ are the same. As a direct result, to investigate the impact of the three factors we discussed in the main text (notching, $\sigma_{\Omega_\mathrm{BNS}^{z^\star}}(f)$, $\delta_\mathrm{Th-CC}(f)$) to the sensitivity of analyses with different $T$ and $\delta f$, we cannot directly compare the values of $\widehat{\sigma}_n^\mathrm{NFC}(f)$ from two analyses, but should consider the fractional increase of $\widehat{\sigma}_n^\mathrm{NFC}(f)$ to the original $\widehat{\sigma}_n(f)$ for the corresponding analysis. We show a direct comparison of this fractional increase for two analyses in Fig.~\ref{fig: 4svs192s}. We stress that Fig.~\ref{fig: 4svs192s} is presented in a logarithmic scale. Although the purple curve exhibits a much higher peak near 15 Hz than the green one, it rapidly drops down to 0 for $f\gtrsim40$ Hz, whereas the green curve maintains an $\mathcal{O}(1)\%$ tail out to very high frequencies. Consequently, the overall sensitivity difference between the two analyses is unlikely to be as dramatic as the purple peak height alone would suggest. A quantitative assessment is left to future work.


\section{What if noise \ac{PSD} is unknown?}\label{app: weighted_avg}
In this final appendix, we discuss the impact of unknown detector noise \ac{PSD} on the analysis. In the previous discussion of this work and our previous work~\cite{Zhong:2022ylh, Zhong:2024dss, Zhong:2025qno} we  neglect an important fact that in practice the detector noise \ac{PSD} is not known \textit{a priori}, and must be inferred from the data. This introduces a subtle but important distinction that, when the estimate \ac{PSD} is no longer constant, the weighted averaging method we use to combine data (c.f. Eq.~\eqref{eq:combine_t}) is no longer equivalent to the simple averaging assumed in the theoretical estimation of the \ac{BNS} foreground. Moreover, we emphasize that we do not have access to \textit{signal-free} data to estimate detector noise \ac{PSD} only. As a result, the estimated detector noise \ac{PSD} can be biased by the presence of underlying \ac{BNS} signals. In this appendix, we aim to roughly quantify the extent of this bias and assess its impact on the analysis, and we leave more thorough investigation for future work. 

To conduct the simulation, now we consider two detectors located at LIGO Hanford and LIGO Livingston, and we inject noise time series into both detectors by \textsc{Bilby} still assuming \ac{CE}2 \ac{PSD}. We randomly choose to consider the Realization \#172 and inject all events with redshift $z>z^\star=0.35$ into the time series. To isolate the effects of using the ground truth of the \ac{PSD} versus the estimated \ac{PSD} on $\widehat{C}(f)$ and $\widehat{\sigma}_n(f)$, we do not perform notching in this test, but focus solely on comparing the resulting $\widehat{C}(f)$ and $\widehat{\sigma}_n(f)$ computed using Eq.~\eqref{eq:stoch} and Eq.~\eqref{eq:combine_t}.
\begin{figure}[!htp]
    \centering
    \includegraphics[width=0.9\linewidth]{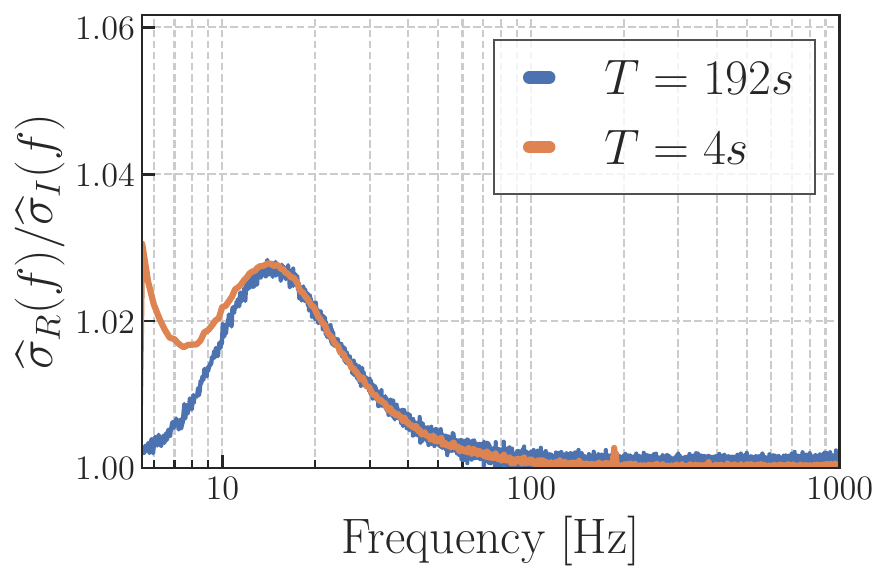}
    \caption{The ratio between $\widehat{\sigma}_R(f)$ and $\widehat{\sigma}_I(f)$, where $\widehat{\sigma}_R(f)$ is calculated using \ac{PSD} estimated via the data, while $\widehat{\sigma}_I(f)$ is computed using the ground truth of the \ac{PSD}.}
    \label{fig: sigmaR_sigmaI}
\end{figure}

The results are shown in Fig.~\ref{fig: Real172}. The left panel corresponds to the $T=192$s case, and the right panel shows the $T=4$s case. In both panels, $\widehat{C}_R(f)$ and $\widehat{\sigma}_R(f)$ are calculated using the estimated \ac{PSD}. Although we also computed $\widehat{C}_I(f)$ and $\widehat{\sigma}_I(f)$ using the injected \ac{PSD}, they are not shown in the figure to maintain visual clarity, as they almost completely overlap with $\widehat{C}_R(f)$ and $\widehat{\sigma}_R(f)$. Instead, we show the absolute difference between $\widehat{C}_I(f)$ and $\widehat{C}_R(f)$ in gray. For the long-segment case ($T=192$ s), this difference is well below the detector noise across the full frequency band. For the short-segment case, the difference remains below $\widehat{\sigma}_R(f)$ in most of the frequency band, but rises to within one to two $\widehat{\sigma}_R(f)$ in the range of $10$Hz $\lesssim f\lesssim20$ Hz.
\begin{figure*}[!htbp]
    \centering
    \includegraphics[width=1.0\linewidth]{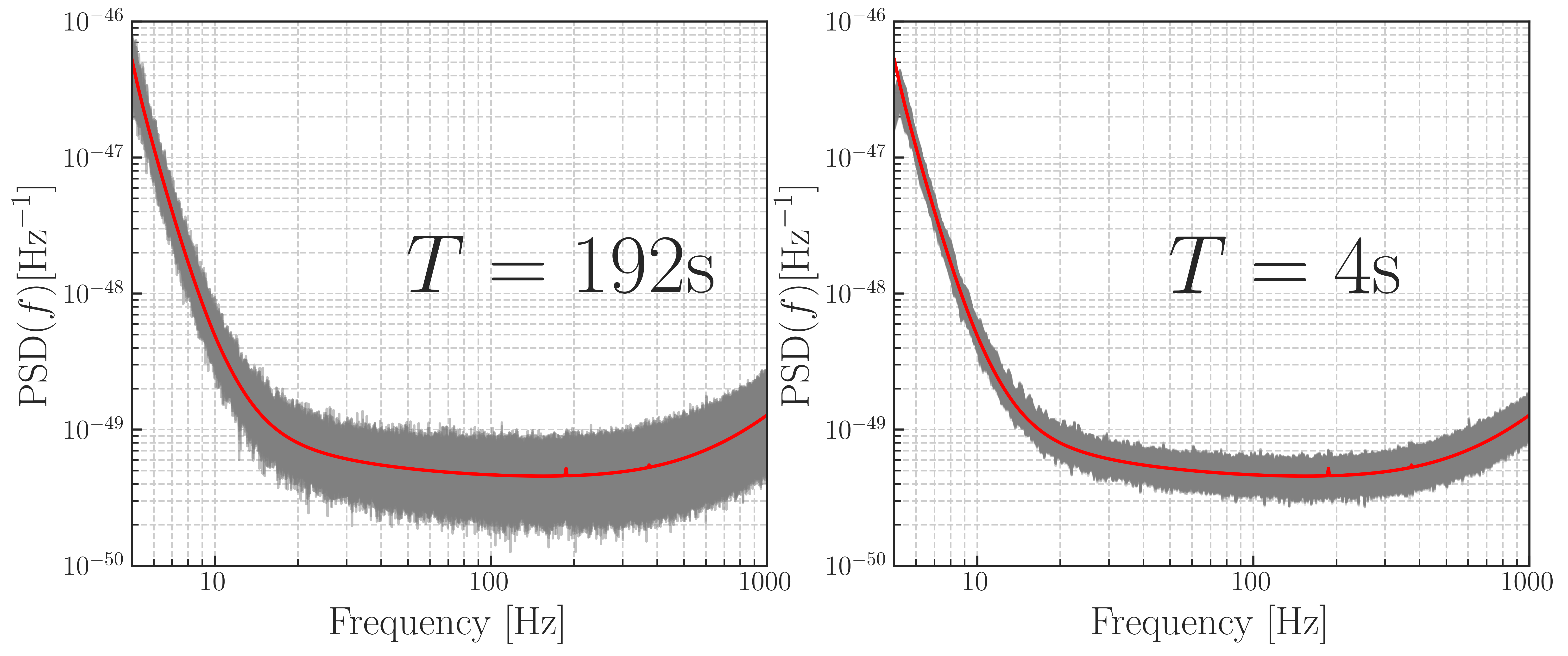}
    \caption{An example of estimated PSD using simulated data. Each gray curve corresponds to an estimate of the noise PSD, and the red curve represents the ground truth.}
    \label{fig: PSD}
\end{figure*}

In Fig.~\ref{fig: sigmaR_sigmaI}, we show the ratio between $\sigma_R(f)$ and $\sigma_I(f)$. As one may expect, the presence of \ac{BNS} signals biases the estimation of the detector noise \ac{PSD}, leading to a slight overestimation in the resulting $\widehat{\sigma}_n(f)$. This bias is often referred to in the literature as confusion noise~\cite{Reali:2022aps,Reali:2023eug}. The confusion noise peaks around $f=14$ Hz, which is close to $f^\star\simeq 13$Hz, where $\sigma_I(f)$ has the minimum value, and then gradually diminishes as $f$ increases, and becomes fully negligible when $f\gtrsim 100$ Hz. We also note that for $T=4$s case, the orange curve has another peak around $f\sim 5$Hz. This feature likely results from the limited frequency resolution of the $T=4$s analysis. Unlike the $T=192$s case, where the signal power is distributed across eight times more frequency bins (1/4 Hz versus 1/32 Hz), the $T=4$ s analysis concentrates signal power into fewer, broader bins, leading to stronger localized bias. 


In Fig.~\ref{fig: PSD}, we display an example of estimated \ac{PSD}s over a stretch of 10,240s long data for both cases. Each gray curve is a individual \ac{PSD} estimate, while the red curve corresponds to the injected \ac{PSD} (ground truth). As shown, without nearby events, the confusion noise is not easily discernible by visual inspection, since there are no clear outliers or systematic deviations in either case.

Taken together, Fig.~\ref{fig: Real172}, Fig.~\ref{fig: sigmaR_sigmaI}, and Fig.~\ref{fig: PSD} demonstrate that, if all the nearby events ($z<z^\star=0.35$) can be perfectly removed, the existence of \ac{BNS} foreground only biases the \ac{PSD} estimation in the frequency range below 100 Hz, and the magnitude of this bias is at most $\lesssim3\%$. Therefore, the resulting overestimation of $\widehat{\sigma}_n(f)$ does not significantly reduce search sensitivity. In fact, this overestimation partially compensates for the other effects discussed in the main text, such as the statistical fluctuation of the foreground and deviations between the measured cross-correlation and theoretical expectation. 
\newpage
\bibliography{references}
\end{document}